\documentclass[compsoc,conference,a4paper,10pt,times]{IEEEtran}
\IEEEoverridecommandlockouts
\usepackage{cite}
\usepackage{amsmath,amssymb,amsfonts}
\usepackage{algorithmic}
\usepackage{graphicx}
\usepackage{textcomp}
\usepackage{bmpsize}
\usepackage{lipsum}
\def\BibTeX{{\rm B\kern-.05em{\sc i\kern-.025em b}\kern-.08em
    T\kern-.1667em\lower.7ex\hbox{E}\kern-.125emX}}

\usepackage[dvipsnames]{xcolor}
\PassOptionsToPackage{table,xcdraw}{xcolor}
\usepackage{booktabs}
\usepackage{multirow}
\usepackage{url}
\usepackage{adjustbox}    
\usepackage{enumitem}
\usepackage{hyperref}

\begin{document}

\title{LFreeDA: Label-Free Drift Adaptation for Windows Malware Detection}


\author{
    \IEEEauthorblockN{Adrian Shuai Li}
    \IEEEauthorblockA{
        Purdue University \\
        Email: li3944@purdue.edu
    }
    \and
    \IEEEauthorblockN{Elisa Bertino}
    \IEEEauthorblockA{
        Purdue University \\
        Email: bertino@purdue.edu
    }
}

\maketitle

\begin{abstract}

Machine learning (ML)-based malware detectors degrade over time as concept drift introduces new and evolving families unseen during training. Retraining is limited by the cost and time of manual labeling or sandbox analysis. Existing approaches mitigate this via drift detection and selective labeling, but fully label-free adaptation remains largely unexplored. Recent self-training methods use a previously trained model to generate pseudo-labels for unlabeled data and then train a new model on these labels. The unlabeled data are used only for inference and do not participate in training the earlier model. We argue that these unlabeled samples still carry valuable information that can be leveraged when incorporated appropriately into training. This paper introduces LFreeDA, an end-to-end framework that adapts malware classifiers to drift without manual labeling or drift detection. LFreeDA first performs unsupervised domain adaptation on malware images, jointly training on labeled and unlabeled samples to infer pseudo-labels and prune noisy ones. It then adapts a classifier on CFG representations using the labeled and selected pseudo-labeled data, leveraging the scalability of images for pseudo-labeling and the richer semantics of CFGs for final adaptation. Evaluations on the real-world MB-24+ dataset show that LFreeDA improves accuracy by up to $12.6\%$ and F1 by $11.1\%$ over no-adaptation lower bounds, and is only $4\%$ and $3.4\%$ below fully supervised upper bounds in accuracy and F1, respectively. It also matches the performance of state-of-the-art methods provided with ground truth labels for $300$ target samples. Additional results on two controlled-drift benchmarks further confirm that LFreeDA maintains malware detection performance as malware evolves without human labeling.

\end{abstract}


\section{Introduction}

Machine learning (ML)-based malware detection systems inevitably face concept drift, where new malware families and evolving variants reduce the effectiveness of models trained on past data~\cite{yang2021cade, chen2023continuous, li2025revisiting}. Maintaining detection accuracy requires periodically updating classifiers using recently collected samples. However, obtaining ground-truth labels for these new samples demands manual analysis or sandbox execution, which is both resource-intensive and slow. In practice, only a very small portion of newly observed samples can be labeled, making label scarcity the central obstacle to sustained malware detection performance.

To reduce human effort, prior research has primarily focused on identifying a small subset of newly observed samples that have drifted from the model’s training data. These samples are then manually labeled or analyzed in sandboxes to obtain ground-truth labels, which are subsequently used for model updates.  The core contributions of these approaches lie in their drift detection (or sample selection) strategies, which aim to identify samples that yield the greatest improvement in model performance under limited labeling budgets. Prior works have explored diverse approaches for drift detection, including uncertainty-based~\cite{chen2023continuous, tripathi2025towards, abusnaina2025exposing}, rejection threshold-based~\cite{barbero2022transcending}, and contrastive learning-driven approaches~\cite{yang2021cade}. While these methods differ in how they detect drifted samples, their retraining procedures remain relatively simple. They rely on either cold-start learning, where a new model is trained from scratch with the newly labeled data, or warm-start learning, which fine-tunes an existing model using those labels~\cite{chen2023continuous}. Recent work~\cite{li2025revisiting} shows that both strategies are suboptimal in low-label settings and introduces an adversarial domain adaptation (AdvDA) method that learns drift-invariant feature representations through joint training on existing and drifted samples. Although these methods reduce labeling costs while maintaining comparable performance, they still require manual annotation and explicit drift detection.

\emph{To address such shortcomings, in this work we propose LFreeDA, an end-to-end label-free drift adaptation framework for Windows malware detection.} LFreeDA adapts malware classifiers to newly collected, unlabeled samples without performing drift detection or requiring expert annotation. The framework jointly trains on existing labeled data and incoming unlabeled data to infer pseudo-labels for the latter, selectively filters them to retain only high-quality pseudo-labels, and then uses the labeled data and the selected pseudo-labeled samples to train a malware classifier in the final stage. Thus, LFreeDA directly leverages the full stream of unlabeled binaries observed in deployment, eliminating the need to identify drifted samples or label them.

A key component of LFreeDA is the use of unsupervised domain adaptation (UDA) to generate pseudo-labels with both high accuracy and sufficient coverage under distribution drift. UDA enables a model to adapt to an unlabeled target domain using labeled data from a related source domain. In the context of concept drift, the source domain corresponds to existing labeled samples, while the target domain consists of newly observed, unlabeled samples. A key requirement of an effective UDA method is to learn feature representations that remain discriminative for the target domain, even when supervision is available only from the source. To meet this requirement, we build on MaxDIRep~\cite{li2023maximal}, which learns representations that are simultaneously discriminative and domain-invariant by decomposing input features into a domain-invariant representation (DIRep) and a domain-dependent representation (DDRep), whose combination reconstructs the original input. To ensure that DIRep preserves information relevant to target labels, MaxDIRep minimizes the information content of DDRep by constraining it with a Kullback-Leibler (KL) divergence to a standard normal distribution. In addition, a GAN-like discriminator enforces domain invariance within DIRep.

However, applying MaxDIRep to malware introduces two challenges not addressed in prior UDA work: (1) class imbalance in real-world PE data, and (2) the inability to directly apply the method to control-flow graph (CFG) representations, which generally provide richer semantics for malware classification~\cite{li2025revisiting}. In real-world scenarios, the majority of PE files are benign, leading to highly imbalanced classes. As shown in our evaluation, when tested on realistic, imbalanced datasets, MaxDIRep exhibits a substantially lower F1-score than accuracy. Furthermore, applying MaxDIRep to CFGs is impractical. The reason is that the model includes an autoencoder component, and state-of-the-art graph autoencoders incur quadratic memory costs and experience steep declines in reconstruction fidelity once graphs exceed a few thousand nodes~\cite{Salha2021FastGAE, Salha2019Degeneracy, Diamant2023Bandwidth}, a common scale for CFGs in real-world binaries.

To address those challenges, LFreeDA (1) introduces a pseudo-label selection mechanism that filters out noisy pseudo-labels using both confidence and class-wise outlier analysis, and (2) deliberately decouples representation roles, using images for scalable pseudo-labeling and CFGs for high-fidelity adaptation, a key design that combines scalability with rich binary-level program semantics.

LFreeDA consists of three main steps. \textit{In Step I (Unsupervised DA)}, we apply MaxDIRep to image representations of labeled source and unlabeled target samples, and use the trained model to predict pseudo-labels for the unlabeled data. \textit{In Step II (High-quality pseudo-label selection)}, we reduce the impact of noisy pseudo-labels by filtering out low-confidence predictions or samples that are distant from their predicted class in the representation space learned from Step I. Prediction probabilities from Step I serve as confidence scores, while outlier detection methods identify samples that deviate substantially from other instances of the same class in the learned representations. \textit{In Step III (Adaptation with selected pseudo-labels)}, we train a classification model using domain adaptation (DA). Our framework supports two strategies: (1) AdvDA~\cite{li2025revisiting}, which trains on existing labeled samples and pseudo-labeled data from Step II; and (2) Fine-tuning (or warm-start training~\cite{chen2023continuous}), which adapts a pre-trained model using pseudo-labeled samples. We evaluate both image- and CFG-based malware representations with these strategies to compare their effectiveness, although a single configuration would be deployed in practice. We reuse the pseudo-labels generated for image representations in Step II as labels for the corresponding CFGs of the same binaries. Our evaluation demonstrates that the additional two steps together improve performance compared to using Step I alone, and that the AdvDA + CFG configuration achieves the best or near-best performance.

We evaluate LFreeDA on a real-world, temporally evolving malware dataset (MB-24+) and two controlled drift benchmarks, BIG-15~\cite{li2025revisiting, ronen2018microsoft} and MalwareDrift~\cite{ma2021comprehensive, wadkar2020detecting}. MB-24+ consists of monthly Windows malware feeds from March-December 2024, with $81-126$ families per month and $27-50\%$ newly observed relative to the previous month, reflecting realistic concept drift in Windows malware. On MB-24+, LFreeDA improves accuracy and F1 by up to $12.6\%$ and $11.1\%$ over no-adaptation baselines, and falls no more than $4\%$ (accuracy) and $3.4\%$ (F1) below fully supervised upper bounds—without using any target labels.

Across all three datasets, we consistently observed that: Step I outperforms all UDA baselines; our pseudo-label selection then further improves pseudo-label quality, yielding on average a $14\%$ accuracy boost while still retaining over $55\%$ of unlabeled samples, and Step III adds a further $4-5\%$ improvement in accuracy/F1 over Step I alone. Compared to AdvDA with a small target-label budget, LFreeDA matches the performance of AdvDA trained with  $50$ labeled target samples on BIG-15 and $300$ on MB-24+, but falls short on MalwareDrift where severe data scarcity in some classes limits pseudo-labeling. We empirically observe that LFreeDA is most effective when each target class has at least $244$ samples, with AdvDA + CFG representations giving the strongest final models.  We also evaluate LFreeDA under obfuscated testing data, where it maintains strong performance with only a modest drop.

\textbf{Contributions.} Our contributions are: 
\begin{itemize}
\item We propose LFreeDA, an end-to-end framework that adapts malware classifiers to concept drift without requiring manual labeling or drift detection. It leverages UDA on malware image representations to generate high-quality pseudo-labels for unlabeled samples, which are then paired with their corresponding image or CFG representations to train the final classifier with DA, with AdvDA + CFG yielding the best performance in our evaluation.

\item We conduct an extensive evaluation of LFreeDA on a real-world dataset with natural drift and two research benchmarks with predefined drift. For each dataset, we assess (1) the contribution of each step to overall performance and (2) the impact of different model configurations within each step.
\item We provide practical deployment recommendations derived from our empirical findings, outlining when LFreeDA is most effective and where its performance may be limited, along with potential strategies for improvement. 
\end{itemize}
\section{Design of LFreeDA}
We first introduce key concepts in Subsection~\ref{problem} and present an overview of the framework in Subsection~\ref{overview}. We then describe its three main stages: (1) training MaxDIRep with labeled source samples and unlabeled target samples (Subsection~\ref{generation}), (2) generating and selecting high-quality pseudo-labels from the unlabeled target samples (Subsection~\ref{selection}), and (3) adapting the model using the selected pseudo-labeled samples (Subsection~\ref{training}).

\subsection{Notions and Definitions}\label{problem}

\subsubsection{Dual representations of a binary} 
LFreeDA uses two representations of each binary $B_i$.

\textbf{Image-based.}  The raw byte stream of $B_i$ is read as a vector of 8-bit unsigned integers and reshaped into a two-dimensional array, yielding $I_i \;=\; (\Phi_i,\; Y_i),$
      where $\Phi_i$ denotes the resulting image matrix and $Y_i\in\{0,1\}^{C}$ is the one-hot encoding of the ground-truth class label (with $C$ classes). Each binary is converted into a grayscale image by mapping consecutive bytes to adjacent pixels, preserving byte-level structure as in \cite{nataraj2011malware}. All images are resized to $56\times56$ using LANCZOS interpolation, matching the configuration in \cite{li2025revisiting}; additional preprocessing details are provided in Appendix~\ref{app_image}.

\textbf{CFG-based.} A CFG is built from the disassembled code of $B_i$.  
      It is represented as $G_i \;=\; (X_i,\; A_i,\; Y_i),$
      where $X_i\in\mathbb{R}^{n_i\times m^{s}}$ is the node-attribute matrix ($n_i$ basic blocks, each with $m^{s}$ source-domain features) and
      $A_i\in\mathbb{R}^{n_i\times n_i}$ is the directed adjacency matrix with $A_i(u,v)$ denoting the control-flow edge from node $u$ to node~$v$. We adopt the same CFG representation as in~\cite{li2025revisiting}, where each node corresponds to a basic block of instructions. For every instruction within a node, we compute an embedding using the PalmTree model~\cite{li2021palmtree}. The node feature is then obtained by averaging these instruction embeddings, and each resulting node feature forms a row in the node-attribute matrix $X_i$.

\subsubsection{Source and target domains} 
We follow the same terminology as prior work, referring to the source and target domains as data collected before and after concept drift, respectively. In real-world deployments, drift is typically assumed to occur gradually over time, as new samples appear and models must be periodically retrained to remain effective. Accordingly, earlier data that have been analyzed and labeled are treated as the source domain, whereas newly collected and unlabeled data represent the target domain. The target domain often includes (1) previously unseen malware families and (2) new variants of existing ones.

Based on this definition, the source domain is fully labeled and provides two modalities derived from each binary:
\[
  \mathcal{I}^{s}=\{(\Phi^{s}_i, Y^{s}_i)\}, 
  \qquad
  \mathcal{G}^{s}=\{(X^{s}_i, A^{s}_i, Y^{s}_i)\}.
\]
The target domain is unlabeled, but similarly provides both image-based and CFG-based representations:
\[
  \mathcal{I}^{t}=\{(\Phi^{t}_i)\}, 
  \qquad
  \mathcal{G}^{t}=\{(X^{t}_i, A^{t}_i)\}.
\]
Ground-truth labels $Y^{t}$ are used only for evaluating the pseudo-label selection module. \textit{Following the setting in~\cite{li2025revisiting}, we focus on the common closed-set setting where the source and target share the same label space.}

\begin{figure}[t]
  \centering
  \includegraphics[width=0.9\linewidth]{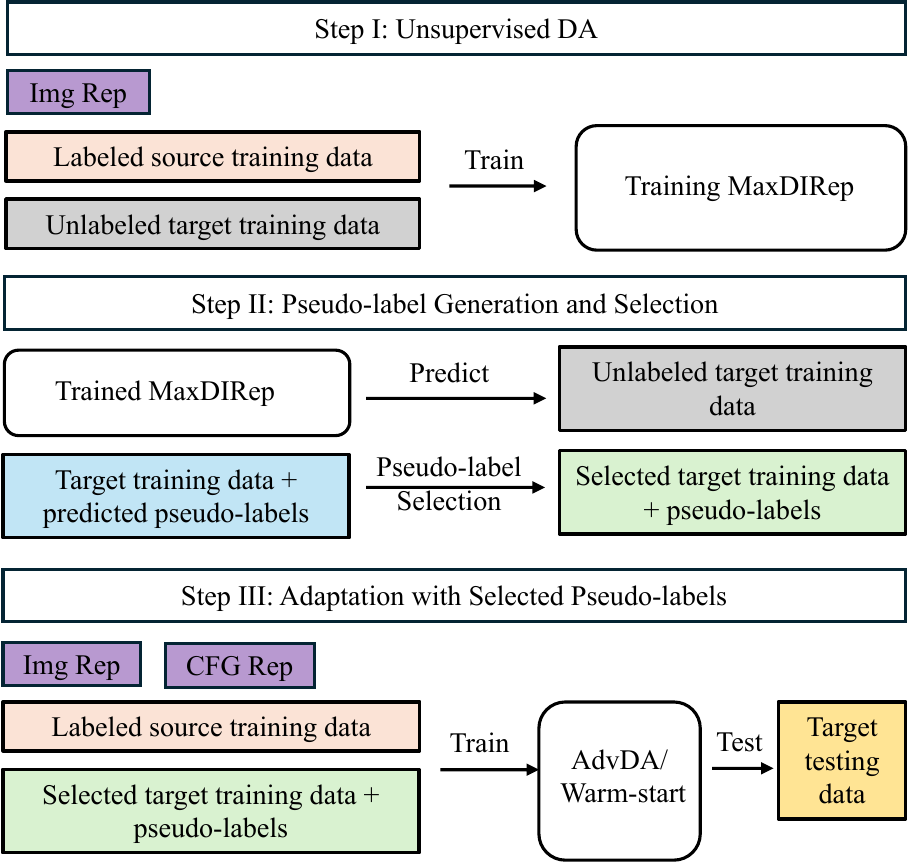}
  \caption{ 
 Overview of the LFreeDA framework. }
  \label{arch}
\end{figure}

\subsection{Overview}\label{overview}
Before presenting an overview of our framework, we introduce an illustrative example to contextualize the problem setting. At the time of model update, we assume access to labeled source data collected before the drift and unlabeled target data collected after the drift. For instance, for a malware detection system retrained monthly, the target training data could consist of binaries collected in the most recent month that analysts have not yet labeled. Samples from earlier months constitute the labeled source training data. The source and target training data are the inputs to LFreeDA, which, after applying the LFreeDA pipeline, outputs a malware classifier used to predict for the future month until the next model update.

Once the source and target training sets are determined, we start the LFreeDA pipeline (Figure~\ref{arch}). Each adaptation process consists of three main stages. (I) We apply MaxDIRep~\cite{li2023maximal} to learn drift-invariant representations from labeled source training and unlabeled target training data using their image-based representations. (II) The trained MaxDIRep then infers pseudo-labels for all target training samples, followed by a pseudo-label selection module that retains only high-quality pseudo-labels per class to mitigate noisy predictions. (III) A new model is trained (distinct from Step~I) using both the labeled source training data and the selected pseudo-labeled target training data. In this stage, we compare two adaptation strategies, AdvDA and fine-tuning. The resulting model is evaluated on the target testing data (e.g., samples from the subsequent month in the above rolling update scenario). In this stage, we also compare image-based and CFG-based representations as input modalities.

\subsection{Step I: Unsupervised DA}\label{generation}

Given labeled source data and unlabeled target data, our goal is to infer accurate pseudo-labels for the latter. DA methods are well-suited for this task, as they transfer knowledge from a source to a related target domain. AdvDA, a leading approach for handling concept drift in malware detection with limited labels, lacks explicit constraints to preserve information useful for target prediction in DIRep, which limits its effectiveness in unsupervised settings—a limitation observed in prior vision studies~\cite{li2023maximal, stojanov2021domain} and confirmed by our experiments across all tested malware datasets.

This occurs because AdvDA’s GAN-like losses only prevent DIRep from revealing domain identity. With a small number of target labels, DIRep can still preserve information useful for predicting those labels while remaining domain-invariant. However, in a fully unsupervised setting with no target labels, there are no constraints preventing AdvDA from discarding information relevant to target prediction, resulting in a DIRep that is domain-invariant but uninformative for the target labels.

To address this limitation, we leverage MaxDIRep~\cite{li2023maximal}, a UDA method originally developed for image classification, as our pseudo-label generation model. MaxDIRep decomposes domain representations into DIRep and DDRep, which are jointly used to reconstruct the input via an autoencoder. Training balances three objectives: a reconstruction loss, ensuring that both representations together preserve sufficient information to reproduce the input; an adversarial domain loss on DIRep, enforcing domain invariance; and a KL-divergence regularization on DDRep, aligning it with a standard normal distribution as a minimal-information baseline. The KL constraint limits DDRep to contain minimal information content, encouraging DIRep to retain features useful for target-label prediction through the reconstruction objective. The adversarial loss further ensures that DIRep remains domain-invariant. As a result, DIRep captures features that are both predictive of target labels and domain-invariant, enabling effective pseudo-label generation.

We train MaxDIRep on the source and target training data using malware image representations as input, while using its original loss functions and training algorithm. The trained model is subsequently used to produce pseudo-labels, prediction probabilities, and intermediate representations for the target training samples. In the following, we describe MaxDIRep’s training objectives and loss functions, and then detail the training procedure and the outputs required by subsequent steps.

\subsubsection{MaxDIRep loss functions}
\begin{figure}[t]
  \centering
  \includegraphics[width=0.9\linewidth]{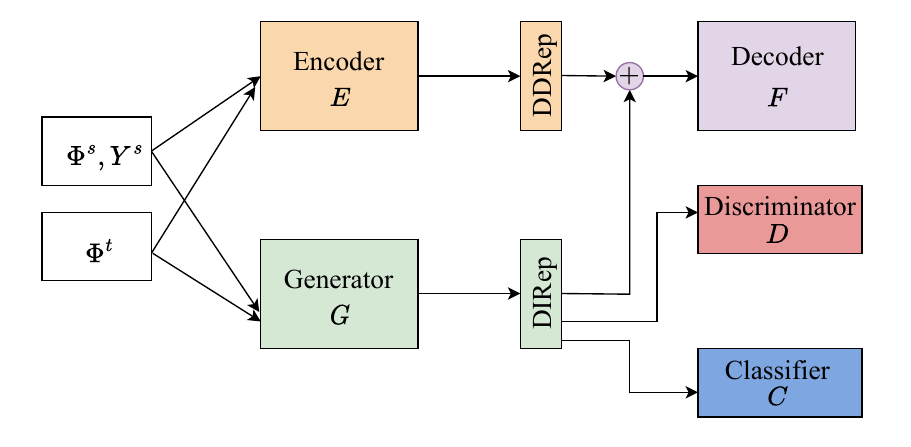}
  \caption{ 
 Architecture of MaxDIRep~\cite{li2023maximal}. }
  \label{maxdirep}
\end{figure}

The source inputs are given by $\mathcal{I}^{s}=\{(\Phi^{s}_i, Y^{s}_i)\}$ and the target inputs are $\mathcal{I}^{t}=\{(\Phi^{t}_i)\}$. Let $G(\Phi;\theta_{g})$ be the generator parameterized by $\theta_{g}$,  which maps an image $\Phi$ to a hidden representation,  $\mathit{DIRep}$, representing features that are common across domains. Let $E(\Phi;\theta_{e})$ be the encoder that maps an image $\Phi$ to a hidden representation $\mathit{DDRep}$, representing domain-specific information. Let $F((G(\Phi) + E(\Phi));\theta_{f})$ be a decoding function mapping the combined hidden representation ($\mathit{DIRep, DDRep}$) to an image reconstruction $\hat\Phi$. Finally, $C((G(\Phi);\theta_{c})$  represents a classifier, parameterized by $\theta_{c}$ that maps from $\mathit{DIRep}$ to the task-specific prediction. The discriminator $D((G(\Phi);\theta_{d})$ maps a $\mathit{DIRep}$ to a prediction of the label $\hat{d} \in \{0,1\}$ of the input sample $\Phi$. The resulting model is shown in Figure~\ref{maxdirep}.

Inference is given by $\hat\Phi = F(G(\Phi) + E (\Phi))$ and $\hat Y = C(G(\Phi))$ where $\hat\Phi$ is the reconstruction of the input $\Phi$ and $\hat Y$ is the prediction. The goal of the training is to minimize the following losses, with respect to their respective parameters:
\begin{align}
    \min_{\theta_g, \theta_e, \theta_f, \theta_c} & \big\{\mathcal{L}_c + \alpha \mathcal{L}_{\text{recon}} + \beta \mathcal{L}_{\text{kl}} + \gamma \mathcal{L}_g \big\} \label{loss_f} \\
    \min_{\theta_d} & \big\{\mathcal{L}_d \big\} \label{eq7}
\end{align}
where $\alpha, \beta, \gamma$ are weights that control the loss terms. The classification loss $\mathcal{L}_c$ trains the model to predict the output labels. Because we assume that the target domain has no labels, the loss applies only to the source domain, and it is defined as follows:
\begin{align}
\mathcal{L}_c=  -  \sum_{i=1}^{N_s} {Y}_i^s \cdot log \hat{Y}_i^s
\end{align}
where $N_s$ represents the number of samples from the source domain, $\hat{Y}_i^s$ is the softmax output of the model.  For the reconstruction loss, we use the mean squared error loss calculated from both domains:
\begin{align}
    \mathcal{L}_{recon} = &\ \sum_{i}^{N_s}||\Phi_i^s - \hat{\Phi}_i^s||_2^2 + \sum_{i}^{N_t}||\Phi_i^t - \hat{\Phi}_i^t||_2^2 
\end{align}%
where $||\cdot||_2^2$ is the squared $L_2$ norm. 

The KL divergence loss $\mathcal{L}_{kl}$ is applied to DDRep to minimize its information content. This encourages the model to rely more heavily on the DIRep for data reconstruction, thereby promoting the retention of task-relevant features in DIRep. Specifically, $\mathcal{L}_{kl}$ measures the divergence between the distribution of DDRep and a standard normal distribution, which serves as a minimal-information baseline. Assuming DDRep follows a Gaussian distribution: $\mathit{DDRep} \sim \mathcal{N}(\mathbb{E}(\mathit{DDRep}), \mathbb{V}(\mathit{DDRep}))$,  the KL divergence loss is defined as:
\begin{IEEEeqnarray}{rCl}
\mathcal{L}_{kl}
&=& D_{KL}(DDRep\,\|\,\mathcal{N}(0,I)) \nonumber\\
&=& -\tfrac{1}{2}\Bigl[\,1+\log\mathbb{V}(DDRep)-\mathbb{V}(DDRep) \nonumber\\
&&\qquad\;\; - \mathbb{E}(DDRep)^{2}\Bigr]
\end{IEEEeqnarray}

The discriminator loss $\mathcal{L}_d$ and generator loss $\mathcal{L}_g$ jointly enforce domain invariance in the learned DIRep representations. The discriminator is trained to predict whether a DIRep originates from the source or target domain, while the generator learns to reduce the discriminator’s ability to distinguish between them. This adversarial process alternates between minimizing the domain prediction loss with respect to $\theta_d$ and maximizing it with respect to $\theta_g$.   $\mathcal{L}_d$ is defined as 
\begin{equation}\label{l_d}
\mathcal{L}_d= 
 -  \sum_{i=1}^{N_s + N_t} \left[ {d}_ilog \hat{d}_i + (1-{d}_i)log (1-\hat{{d}}_i) \right] 
\end{equation}%
where $d_i \in \{0,1\}$ is the ground truth domain label for samples $i$. $\mathcal{L}_g$  is defined with inverted ground truth domain labels, as we want to minimize the domain prediction accuracy: 
\begin{equation}\label{l_g}
\mathcal{L}_g= 
 -  \sum_{i=1}^{N_s + N_t} \left[ (1-{d}_i)log \hat{d}_i + d_ilog (1-\hat{{d}}_i) \right]
\end{equation}%

\subsubsection{Training MaxDIRep}Training of the generator $G$ jointly optimizes $\mathcal{L}_g$, $\mathcal{L}_c$, and $\mathcal{L}_{\text{recon}}$, aiming to minimize domain classification accuracy while maximizing label prediction accuracy and reconstruction quality. The discriminator is trained with $\mathcal{L}_d$ to maximize domain classification accuracy, and the classifier is trained with $\mathcal{L}_c$. The encoder-decoder pair $(E, F)$ functions similarly to a variational autoencoder, except that the decoder $F$ reconstructs inputs from the combined DIRep and DDRep. The encoder $E$ minimizes $\mathcal{L}_{\text{kl}}$ and $\mathcal{L}_{\text{recon}}$, while $F$ minimizes $\mathcal{L}_{\text{recon}}$. Training proceeds with mini-batch gradient descent in alternating steps: first updating $G$, $E$, $F$, and $C$ to minimize their respective losses while freezing the discriminator, then updating the discriminator to minimize $\mathcal{L}_d$. Upon convergence, DIRep becomes discriminative for class prediction yet invariant to domain. We use the same loss-coefficient settings as prior work (see Appendix~\ref{app_mb_DA}).

\subsubsection{Inference on the target training set}
After training MaxDIRep, we extract the following outputs for the target training set ${\Phi^t}$: (1) predicted labels $\hat{Y}^t$; (2) prediction probabilities ${P}^t$ from the softmax layer of the classifier $C$; and (3) domain-invariant representations $G(\Phi^t)$.

\subsection{Step II: Pseudo-label Selection}\label{selection}

\subsubsection{Motivating example}

\begin{figure}[t]
  \centering
  \includegraphics[width=0.9\linewidth]{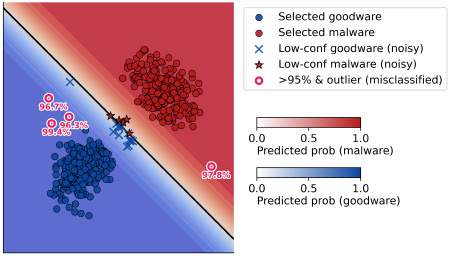}
  \caption{ 
Illustration of the class-wise outlier and confidence-based pseudo-label selection (best viewed in color). The line is the decision boundary.   }
  \label{select}
\end{figure}

Our approach to pseudo-label selection is guided by two principles. First, samples located near the decision boundary are more susceptible to misclassification due to model uncertainty. Second, samples that appear as outliers within their predicted class are also likely to be incorrectly labeled. This concept is illustrated in Figure~\ref{select}, where the misclassified red circles and the low-confidence noisy data points can be identified using these two criteria. The selected samples in each class form a cluster and are further away from the frontier. 

To identify samples near the decision boundary, a common strategy is to use prediction confidence, under the common assumption that higher confidence corresponds to greater distance from the boundary~\cite{kan2021investigating}. However, confidence alone is insufficient, as incorrect predictions may still have high certainty. To address this limitation, we introduce an additional selection step: for each predicted class, we apply an outlier detection algorithm to identify samples that, despite high confidence, deviate significantly from the distribution of their assigned class. This further helps detect mislabeled instances, such as those labeled ``$\mathit{>95\%\ \&\ outlier}$'' in Figure~\ref{select}.

Prior works have shown that prediction confidence~\cite{hendrycks2016baseline, abusnaina2025exposing, tripathi2025towards} and outlier detection~\cite{abusnaina2025exposing} are effective for detecting concept drift by identifying samples that have drifted. In contrast, we adapt these techniques to the problem of pseudo-label selection, using confidence and outlier scores to filter out noisy pseudo-labels. These methods are well-suited for this task because, at a high level, both drift detection and pseudo-label selection share the same underlying goal: identifying samples that deviate from their predicted class.

\subsubsection{Details on the pseudo-label selection}
To implement the above selection strategies, we utilize the inference results on the target training set, including predicted labels $\hat{Y}^t$, prediction probabilities ${P}^t$ from the softmax layer of classifier $C$, and domain-invariant representations $G(\Phi^t)$. We apply an outlier detection algorithm separately to each group of samples sharing the same predicted class. The algorithm operates on the DIReps of these samples to identify inliers and outliers within each class. This offers a key advantage over input-space outlier detection by enabling predictions to be validated directly in the latent space, where class boundaries are more clearly defined. Finally, we select a sample $i$ and its pseudo-label $\hat{Y}^t_i$, only if they satisfy \textbf{two conditions:} (1) The prediction probability  ${P}^t_i \geq 0.95$, as prediction below the threshold is considered as low-confidence; (2) sample $i$ is an inlier within its own predicted label $\hat{Y}^t_i$. As we will show in the experiments, these two conditions combined are better than using the confidence-based filtering alone.

Since outlier detection plays a critical role in the accuracy of our pseudo-label selection module, we evaluate five candidate algorithms: Local Outlier Factor (LOF)~\cite{Breunig2000}, Gaussian Mixture Model (GMM)~\cite{Bishop2006}, One-Class Support Vector Machine (OC-SVM)~\cite{Scholkopf2001}, Mahalanobis Distance-based Detector (Mahal-D)~\cite{RousseeuwHubert2018}, and Isolation Forest (iForest)~\cite{Liu2008IsolationForest}. Details of each method and how they are integrated are included in Appendix~\ref{app_ODA}.

\subsubsection{Conservative pseudo-label filtering}
Step II intentionally adopts a conservative filtering strategy that tends to underestimate pseudo-label quality. Instead of relying on confidence only, we require each sample to have a high confidence score and be an inlier within its predicted class in the DIRep space. This additional constraint filters out high-confidence yet misclassified samples by validating predictions through the model’s internal representations. For example, a sample predicted as class A with 97\% confidence but actually belonging to class B will appear anomalous among true A samples and thus be excluded.

Although this approach may reject some correctly labeled samples, the trade-off is justified. The greater risk lies in retaining mislabeled, high-confidence samples, which can severely degrade Step III performance. Since AdvDA adapts well with limited target labels~\cite{li2025revisiting}, a smaller, higher-quality pseudo-labeled set yields better adaptation than a larger, noisier one. In the following, we introduce a metric designed to jointly assess pseudo-label quality and quantity. In the evaluation sections, we will demonstrate that our conservative selection still preserves a substantial portion of reliable pseudo-labels.

\subsubsection{Evaluation metrics for pseudo-label selection}
To evaluate the performance of our pseudo-label selection method and compare different outlier detection algorithms, we use two key metrics. The first is \textit{pseudo-label accuracy ($a$)}, which measures the percentage of retained samples whose pseudo-labels match their ground-truth labels. True labels are used strictly for evaluation and are not involved in pseudo-label generation.

The second metric is \textit{coverage ($c$)}, defined as the fraction of kept target training samples. Since both $a$ and $c$ are bounded within the interval $[0,1]$, we combine them without normalization to compute the \textit{Accuracy-Coverage Score (ACS)}, defined as:
\begin{equation} ACS = w \times a + (1 - w) \times c \end{equation}
where $w$ is a weighting factor, set to $0.5$ in our experiments to give equal importance to accuracy and coverage. The ACS provides a balanced measure of performance, reflecting both the quality and quantity of the selected pseudo-labels, capturing the ability to keep a large fraction of samples while maintaining high labeling accuracy.


\subsection {Step III: Adaptation with Pseudo-labels}\label{training}

In Step III, DA is performed using both the labeled source data and the pseudo-labeled target data selected in Step II. The current implementation of the framework supports two DA strategies: (1) AdvDA - a state-of-the-art adaptation method designed to mitigate concept drift in malware detection~\cite{li2025revisiting}; (2) Warm-start training, wherein a model trained with source data is fine-tuned using the pseudo-labeled target samples. Furthermore, each adaptation method is trained on both CFG and image representations. In the following section, we present a summary of the adaptation models trained and evaluated in Step III.

 Let the final training data in image representation be $\bigl(\mathcal{I}^{s},\;\hat{\mathcal{I}}^{t}_{\mathrm{high}}\bigr)$ where $\mathcal{I}^{s}$ is the source image data and  $\hat{\mathcal{I}}^{t}_{\mathrm{high}}$ is the target image data with high-quality pseudo-labels. For CFG, the final training data is  $\bigl(\mathcal{G}^{s},\;\hat{\mathcal{G}}^{t}_{\mathrm{high}}\bigr),$ where $
  \hat{\mathcal{G}}^{t}_{\mathrm{high}}
  =\{(X^{t}_i, A^{t}_i, \hat{Y}^{t}_i)\mid(\Phi^{t}_i,\hat{Y}^{t}_i)\in\hat{\mathcal{I}}^{t}_{\mathrm{high}}\}.
$  Since malware image representations are used in the first two steps, 
we extract the corresponding CFGs for the samples in $\hat{\mathcal{I}}^{t}_{\mathrm{high}}$. With the final training datasets for both representations in place, we next describe the adaptation models used in Step III.

 \textbf{Warm-start GIN:} We train a Graph Isomorphism Network (GIN) with Jumping Knowledge~\cite{xu2018powerful, li2025revisiting} using the source domain graph data $\mathcal{G}^{s}$, and subsequently fine tune it with the target graph data $\hat{\mathcal{G}}^{t}_{\mathrm{high}}$. GIN effectively captures graph structural information by aggregating features from adjacent nodes and generating graph-level representations through a pooling mechanism. Empirical results demonstrate that GIN outperforms other graph neural networks (GNN) in the malware detection task~\cite{wu2021malware}.

 \textbf{AdvDA + GIN~\cite{li2025revisiting}}: This method is trained using both the source graph data $\mathcal{G}^{s}$ and target graph data $\hat{\mathcal{G}}^{t}_{\mathrm{high}}$. It is the state-of-the-art adaptation approach assuming limited labeled data in the target domain. Similar to MaxDIRep, AdvDA learns DIReps for training a classifier, but the two methods differ in that:
\begin{itemize}
\item Architecture: AdvDA does not have the upper network components present in MaxDIRep (Figure~\ref{maxdirep}), namely, Encoder, DDRep, and Decoder, as well as reconstruction loss $\mathcal{L}_{\text{recon}}$ and KL divergence loss $\mathcal{L}_{\text{kl}}$.
\item Classification loss: The classification loss in AdvDA is applied to both source and target domains. 
\item Graph-based input: Unlike MaxDIRep, which uses a CNN-based generator, AdvDA employs a GIN-based generator and is originally designed to operate on CFGs from both domains.
\end{itemize}
The core idea behind AdvDA + GIN is to learn DIReps from source and target CFGs that are structurally and semantically similar. This allows a classifier trained on source graphs together with a small set of labeled target graphs to generalize more effectively to the target testing data. To achieve this, AdvDA employs the same domain adversarial training as MaxDIRep, where the generator is trained to maximize the discriminator’s loss using the adversarial objectives defined in~(\ref{l_d}) and~(\ref{l_g}).

 \textbf{Warm-start (ResNet-50):} This follows standard practice in malware image classification under concept drift~\cite{ma2021comprehensive, vasan2020imcfn, bhodia2019transfer, kumar2021mcft}. We initialize the ResNet-50 model with weights pre-trained on the ImageNet dataset~\cite{deng2009imagenet}, leveraging its ability to extract general visual features. The model is then fine-tuned using both the source domain image data $\mathcal{I}^{s}$ and target domain image data $\hat{\mathcal{I}}^{t}_{\mathrm{high}}$. This warm-start strategy enables the model to retain knowledge of general images while adapting to domain-specific characteristics of malware images, thereby enhancing its robustness to distributional shifts.

\textbf{AdvDA + CNN}: This variant of AdvDA operates on image-based representations of malware rather than graph-based inputs. We adopt the original AdvDA + CNN architecture from~\cite{li2025revisiting}, which uses a multi-layer convolutional neural network (CNN) for the generator. The loss functions and training procedure follow the same design as the graph-based AdvDA.

While LFreeDA supports four model configurations based on two adaptation strategies and two malware representations, only one would typically be deployed in practice. We evaluate all four to compare their performance and guide deployment decisions.

\section{Evaluation: Natural Drift on Real-World Malware} \label{evaluation:mb}
We begin by evaluating LFreeDA on a real-world malware dataset collected over time, adopting a practical monthly rolling update where each model is tested on data from the following month, reflecting natural concept drift in malware evolution. This setup illustrates how LFreeDA can be deployed in practice. We further evaluate LFreeDA on two benchmark datasets with predefined drift, presented in Sections~\ref{eva:big15} and~\ref{evaluation:family}.

\subsection{Dataset and Experiment Setup}

\subsubsection{Dataset} Existing real-world Windows malware datasets are largely outdated and often lack original binaries. BODMAS~\cite{bodmas} and EMBER~\cite{2018arXiv180404637A} are two widely used longitudinal datasets, but neither meets the requirements of our study. Both provide extracted features for malware and benign samples; however, BODMAS includes only malware binaries collected over five years ago, while EMBER was released seven years ago and does not include any binaries. The original EMBER samples are available only through VirusTotal, which requires an enterprise subscription, making them inaccessible for public use.

To overcome these limitations, we use an extended version of the MB-24 dataset~\cite{li2025revisiting}, originally comprising malware binaries from the MalwareBazaar daily feed~\cite{malwarebazaar_api} collected between March and September 2024. We add samples from October to December 2024 and refer to the resulting dataset as MB-24+. Each month in MB-24+ contains at least $81$ distinct malware families, with $27\%$-$50\%$ newly observed relative to the previous month. We verified that no duplicate binaries (based on SHA-256 hashes) exist within or across months. Overall, MB-24+ provides a more realistic reflection of the long-term temporal evolution and distributional drift of real-world malware, offering a challenging and representative benchmark for evaluating LFreeDA. Detailed monthly statistics of MB-24+ are reported in Table~\ref{newmb24}.


\begin{table}[h]
\caption{Monthly statistics of the MB-24+ dataset, including the total number of samples, total malware families, number of newly observed (unseen) families compared to the previous month, and the unseen family ratio (unseen families / total families). March 2024 is the initial month and thus has no unseen families or ratio.}
\label{newmb24}
\centering
\resizebox{0.85\linewidth}{!}{
\begin{tabular}{@{}crrrr@{}}
\toprule
\textbf{Month/Year} & \multicolumn{1}{c}{\textbf{\begin{tabular}[c]{@{}c@{}}Total \\ Samples\end{tabular}}} & \multicolumn{1}{c}{\textbf{\begin{tabular}[c]{@{}c@{}}Total\\  Families\end{tabular}}} & \multicolumn{1}{c}{\textbf{\begin{tabular}[c]{@{}c@{}}New/Unseen \\ Families\end{tabular}}} & \multicolumn{1}{c}{\textbf{\begin{tabular}[c]{@{}c@{}}Unseen \\ Ratio (\%)\end{tabular}}} \\ \midrule
03/2024             & 1505                                                                                  & 105                                                                                    & -                                                                                           & -                                                                                        \\
04/2024             & 1080                                                                                  & 81                                                                                     & 22                                                                                          & 27                                                                                    \\
05/2024             & 1496                                                                                  & 100                                                                                    & 44                                                                                          & 44                                                                                     \\
07/2024             & 1618                                                                                  & 126                                                                                    & 60                                                                                          & 48                                                                                    \\
08/2024             & 1613                                                                                  & 114                                                                                    & 46                                                                                          & 40                                                                                    \\
09/2024             & 1337                                                                                  & 92                                                                                     & 32                                                                                          & 35                                                                                    \\
10/2024             & 1444                                                                                  & 97                                                                                     & 45                                                                                          & 46                                                                                    \\
11/2024             & 1210                                                                                  & 108                                                                                    & 54                                                                                          & 50                                                                                    \\
12/2024             & 1302                                                                                  & 105                                                                                    & 46                                                                                          & 44                                                                                    \\ \bottomrule
\end{tabular}
}
\end{table}

For the benign class, we use the same dataset as~\cite{li2025revisiting}, which was constructed by collecting Windows PE files from virtual machines configured with clean installations of Windows 8, 10, and 11, and common Windows applications. This process yielded $\text{16,000}$ benign binaries.

\subsubsection{Source and target dataset preparation}\label{mb_setup}
We adopt a time-consistent data split strategy following~\cite{li2025revisiting, chen2023continuous}. Malware samples from March-May 2024 constitute the source domain (pre-drift data), which is divided into 75\% for source training and 25\% for source testing. Starting in July 2024, each month’s data is used for the new target-domain training, and the trained LFreeDA is subsequently evaluated on the following month. This rolling-adaptation procedure continues until November 2024, covering the post-drift period (July-December 2024). We skip June to ensure a clean temporal separation between pre- and post-drift periods. The resulting sequence of model adaptation is July $\rightarrow$ August, August $\rightarrow$ September, September $\rightarrow$ October, October $\rightarrow$ November, November $\rightarrow$ December, where each pair $(\mathit{target\ training\ month} \rightarrow \mathit{testing\ month})$ shares the same source data used to train  LFreeDA. Compared to~\cite{li2025revisiting}, which considered only two updates (July $\rightarrow$ August, August $\rightarrow$ September), our setup includes three additional updates, providing a more comprehensive longitudinal evaluation of LFreeDA.

Since benign PE files lack temporal labels, we follow the same setup as in~\cite{li2025revisiting}, which assumes that the benign distribution remains relatively stable over time. In their configuration, $8{,}000$ benign samples are used for the source domain, and the remaining $8{,}000$ are evenly divided between target training and target testing. Using the same partitioning strategy and CFG extraction pipeline, we successfully extracted $6{,}510$ benign CFGs for the source and $5{,}768$ for the target, consistent with the results reported in~\cite{li2025revisiting}. Extraction failures for benign PE files occurred due to IDA Pro disassembly errors during CFG construction, while all malware samples were successfully processed.  We next report the malware-to-benign ratios in the CFG- and image-based representations for both the source and target datasets. 

For the CFG-based representation, the source training and testing sets maintain a malware-to-benign ratio of $0.6{:}1$. Because the number of malware samples varies slightly across post-drift months, this ratio cannot be perfectly preserved in every target training and testing set; however, it remains close, with an average ratio of approximately $0.5{:}1$ across all tasks. The exact ratios for the source, target-training, and target-testing partitions in each testing month are provided in Table~\ref{mb24_ratio} (Appendix). This configuration avoids spatial bias~\cite{pendlebury2019tesseract} by ensuring that the target training ratio remains realistic and reflective of the target testing distribution.

For the image-based representation, the source domain maintains a malware-to-benign ratio of $0.5{:}1$, while the target sets exhibit a stable ratio of roughly $0.3{:}1$. Because image representations can be extracted for all benign PE files, the absolute ratios are lower than those in the graph-based setup. Nonetheless, each modality maintains internally consistent ratios between target training and testing sets across all tasks (see Table~\ref{mb24_ratio} for details).

\subsection{Evaluation Results}

LFreeDA consists of three main steps, and we evaluate each step separately to highlight its individual contribution. Subsections~\ref{mb24:step1} and~\ref{mb24:step2} present the results of Steps~I and~II, respectively, together with comparisons to alternative or baseline methods. The final results after all three steps are summarized in Subsection~\ref{mb24:step3}.

\begin{figure}[t]
  \centering
  \includegraphics[width=0.75\linewidth]{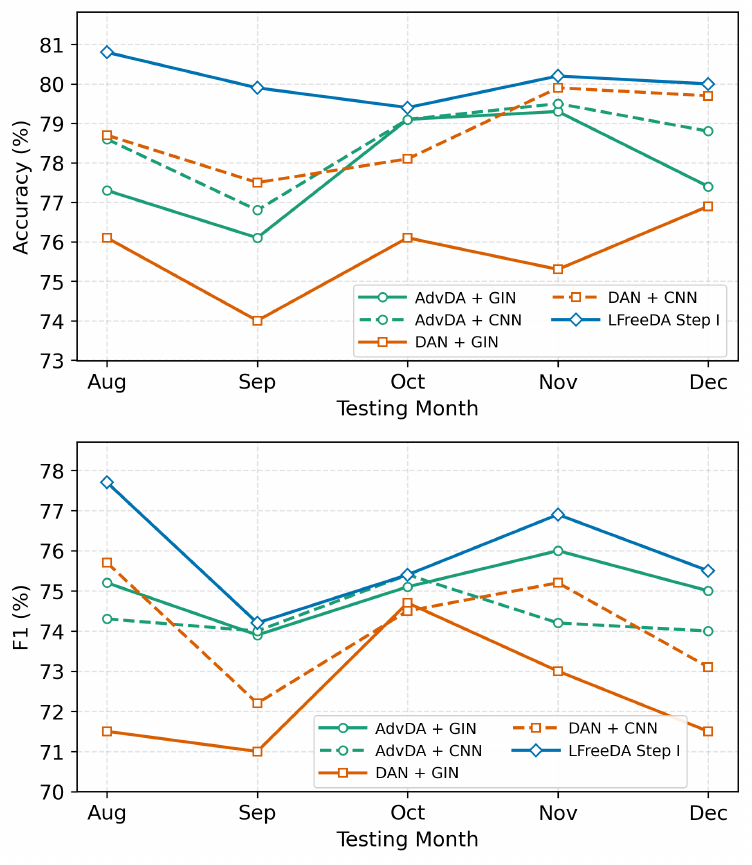}
  \caption{Accuracy (top) and F1 (bottom) after Step~I of LFreeDA and other UDA baselines across post-drift testing months on the MB-24+ dataset.}
\label{mb24-step1}
\end{figure}

\subsubsection{Evaluation of Step I}\label{mb24:step1}


We evaluate Step I of LFreeDA against established DA methods designed for concept drift, including AdvDA + GIN, AdvDA + CNN, DAN + GIN, and DAN + CNN. DAN~\cite{long2015learning} employs Maximum Mean Discrepancy (MMD) loss, a kernel-based distance metric, to align the distributions of source and target hidden representations. These baselines have demonstrated state-of-the-art performance in Windows malware detection when $20$-$500$ labeled target samples are available for training~\cite{li2025revisiting}. However, they have not been tested in a fully unsupervised scenario where the target training data contains no labels. When we refer to AdvDA in Step I, it is a purely unsupervised baseline trained only on source labels. In Step III, AdvDA is the final adaptation method trained on source labels plus pseudo-labels from Step II.  Implementation details for all methods are provided in Appendix~\ref{app_mb_DA}.

Each model is trained using the labeled source dataset and the unlabeled target training set, and evaluated on the target testing set. Performance is measured by accuracy and macro-F1, and all results are averaged over five independent runs. Figure~\ref{mb24-step1} shows the longitudinal accuracy and F1 trends across the post-drift testing period. Step~I of LFreeDA consistently outperforms all four baselines. Between AdvDA and DAN, AdvDA shows overall better performance regardless of the representation used.  DAN performs better on the image-based representation but shows reduced effectiveness on the CFGs.

\begin{figure}[t]
  \centering
  \includegraphics[width=\linewidth]{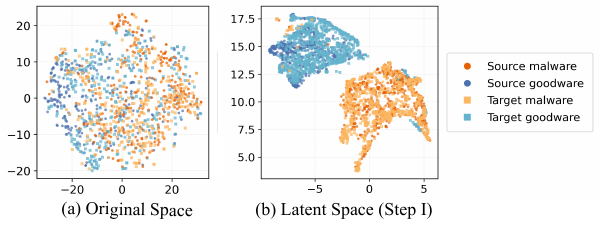}
  \caption{T-SNE visualization of the original feature space and the latent space of LFreeDA Step I (DIRep features) for the Jul $\rightarrow$ Aug adaptation task. Visualizations for all the tasks are provided in Figure \ref{app_mb24_tsne_full}.}
  \label{mb24_tsne}
\end{figure}

We provide further evidence that the DIRep features extracted in Step I are domain-invariant, that is, the DIRep alone does not reveal whether a sample originates from the source or target domain. To demonstrate this, we used the trained Step I model to extract DIReps from both source and target training inputs. Figure~\ref{mb24_tsne} shows t-SNE visualizations comparing the learned latent space with the original feature space. Step~I effectively reduces the distribution divergence between source and target data, as samples from the same class but different domains are closely aligned. The clear class boundaries also demonstrate that the learned representations remain highly discriminative. Similar patterns are observed across all other model adaptation tasks (see Figure~\ref{app_mb24_tsne_full}).

\begin{figure}[t]
  \centering
  \includegraphics[width=0.9\linewidth]{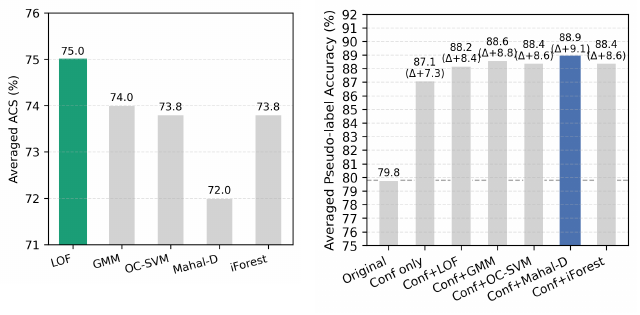}
  \caption{ The left graph shows the averaged ACS scores for each outlier detection method combined with confidence filtering on MB-24+. The right graph
compares averaged pseudo-label accuracy: original, with confidence filtering, with
confidence filtering + outlier detection.}
  \label{mb24_step2}
\end{figure}

\begin{figure*}[t]
  \centering
  \includegraphics[width=0.85\linewidth]{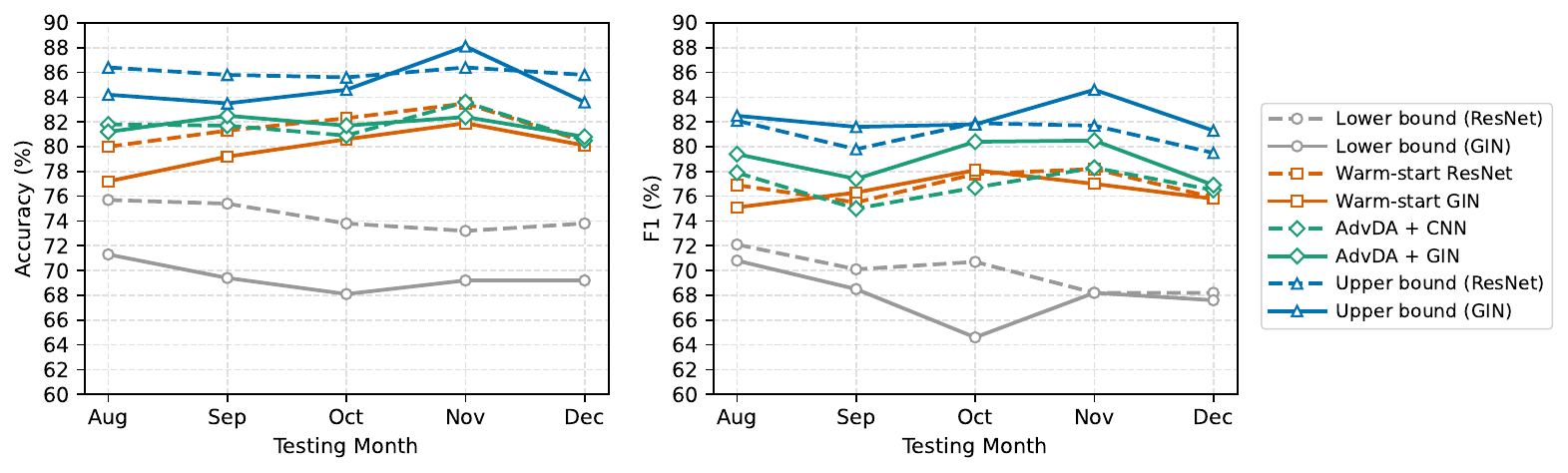}
  \caption{Accuracy (left) and F1 (right) after Step~III of LFreeDA across post-drift testing months on the MB-24+ dataset. The results are compared with the lower and upper bound baselines.}\label{mb-step3}
\end{figure*}

\subsubsection{Evaluation of Step II}\label{mb24:step2}


To evaluate the effectiveness of pseudo-label selection, we perform the analysis separately for each model update. In every update (e.g., July $\rightarrow$ August, August $\rightarrow$ September, and so on), the Step~I model trained for that specific update is applied to its corresponding target-training set to obtain predicted labels, confidence scores, and DIRep features. These outputs, together with the ground-truth labels (used only for evaluation), are used to assess pseudo-label selection performance. For consistency, we fix the confidence threshold at $0.95$ and mark $20\%$ of samples as outliers across all five outlier-detection algorithms (implementation details in Appendix~\ref{app_mb_out}).

For each adaptation task, we measure (1) the initial prediction accuracy from Step~I, (2) the accuracy after confidence-based filtering, and (3) the accuracy after applying both confidence-based filtering and outlier detection. Figure~\ref{mb24_step2} presents results averaged across all model updates: the left panel reports the Accuracy-Coverage Score (ACS) for each outlier detection method used in conjunction with confidence-based filtering, while the right panel shows the corresponding averaged pseudo-label accuracy (excluding coverage). Detailed per-update statistics for accuracy and coverage are provided in Table~\ref{app:mb_outlier_stats}.

We observe that:

\begin{itemize}
    \item LOF achieves the highest ACS across all target tasks, keeping an average of $61\%$ of samples while achieving an average labeling accuracy of $88.2\%$. This represents an $8.4\%$ improvement over the initial average pseudo-label accuracy of $79.8\%$. Although Mahal-D attains the highest average labeling accuracy, its ACS is the lowest due to limited coverage. LOF offers a better balance between the quality and quantity of selected pseudo-labels, which is key to the successful adaptation in Step~III.
    \item Ablation experiments confirm the effectiveness of our pseudo-label selection strategy, showing that the combination of confidence-based filtering and class-wise outlier detection yields the highest pseudo-label accuracy.
\end{itemize}
These selected high-quality pseudo-labels are then used in Step~III for model adaptation.

\subsubsection{Evaluation of Step III (Final Results)}\label{mb24:step3}
We train the four adaptation models described in Section~\ref{training} with the rolling adaptation setup introduced earlier. In each adaptation step, the models have access to the source training data with ground truth labels and the current target training month data with the selected pseudo labels. The adapted models are then evaluated on the data from the following month, which serves as the testing period. For each modality, we also report \textit{lower bound} and \textit{upper bound} baselines. The lower bound represents a no-adaptation setting, where a model trained solely on the source data is directly evaluated on each testing month without any updates. In contrast, the upper bound is obtained by fine-tuning the same pretrained source model on each target training set using ground truth labels, assuming all target samples are labeled, and then evaluating it on the corresponding testing data. For the image-based representation, we employ ResNet50 as the base model for both bounds, while for the CFG-based representation, we use GIN. Implementation details are provided in Appendix~\ref{app_mb_train}.

Note that pseudo-label selection is performed after Step~I, which uses the image representations of binaries. In cases where a corresponding CFG representation is unavailable for a selected pseudo-labeled sample, that sample is excluded from training. Performance is measured by accuracy and macro-F1, and all results are averaged over five independent runs. Figure~\ref{mb-step3} shows the longitudinal accuracy and F1 trends across the post-drift testing period. Our observations are as follows:

\begin{itemize}
    \item AdvDA + GIN achieves the highest overall performance in terms of both accuracy and F1 across all test months, with an average accuracy of $82\%$ and an average F1 of $79\%$. This corresponds to an improvement of $7.6$-$12.6\%$ in averaged accuracy and $9.1$-$11.1\%$ in averaged F1 compared to the two lower bounds, while remaining only a small margin below the two upper bounds (a gap of $2.8$-$4\%$ in accuracy and $2$-$3.4\%$ in F1).
    \item The best observed average accuracy of $82\%$ is comparable to that reported in~\cite{li2025revisiting}, where AdvDA was trained with $300$ labeled target samples. This demonstrates that LFreeDA achieves similar performance without using any labeled target data. Moreover, this represents a conservative comparison, since~\cite{li2025revisiting} evaluated only two model updates over a three-month drift period, whereas our setup includes three additional updates spanning six months in total.
    \item The CFG-based representation also shows a clear F1 gain when using AdvDA, consistent with the findings in~\cite{li2025revisiting}. However, this advantage is not observed for the warm-start training.
\end{itemize}

\subsection{Impact of Obfuscation}\label{evaluation:obf}
\begin{figure}[t]
  \centering
  \includegraphics[width=0.95\linewidth]{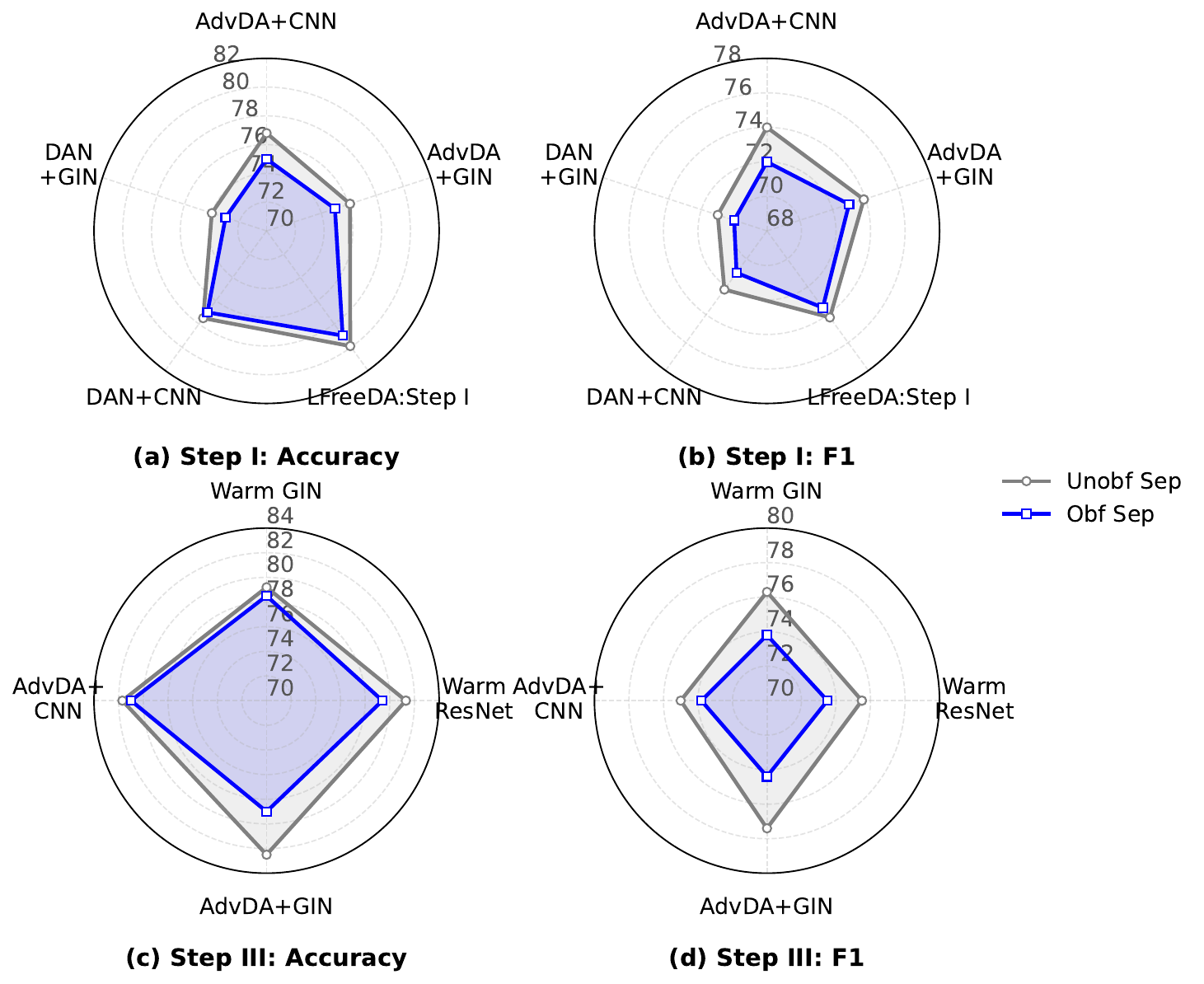}
  \caption{Performance of LFreeDA on the obfuscated September testing data. (a) Accuracy after Step~I, (b) F1 after Step~I, (c) Accuracy after Step~III, and (d) F1 after Step~III.}\label{obf}
\end{figure}

To evaluate robustness against obfuscated malware, we design an experiment using the August $\rightarrow$ September task. The malware samples in the source training data (March-May) and target training data (August) are unobfuscated, while the target testing set (September) contains obfuscated malware generated with Hyperion \cite{nullsecurity_binary_tools}, a runtime PE crypter. This setup captures a challenging scenario where models trained only on unobfuscated malware are tested on unseen obfuscated malware, providing a lower bound on LFreeDA’s robustness when new packing or encryption techniques appear without prior examples. Prior work~\cite{li2025revisiting} shows that adding obfuscated samples during training further improves generalization to obfuscated variants, and similar gains can be expected for LFreeDA as well. In this work, we focus on this lower-bound setting by evaluating Step~I of LFreeDA and all baselines from Section~\ref{mb24:step1}, as well as Step~III of LFreeDA, on the obfuscated September test set, with results reported in Figure~\ref{obf}.
 
LFreeDA exhibits strong robustness to obfuscation. Step~I maintains stable performance, while Step~III shows only a minor drop in accuracy and F1, even though the training data in this experiment were intentionally kept unobfuscated to evaluate robustness. On the obfuscated September test set, Step III accuracy decreases by approximately $1$-$3.5\%$ and F1 by $1$-$3.0\%$. The largest decline occurs with AdvDA + GIN (about $3.5\%$ in accuracy and $3.0\%$ in F1), whereas AdvDA + CNN remains the most stable, with only around a $1\%$ drop in both metrics.

\section{Evaluation: Controlled Drift on BIG-15 Benchmark}\label{eva:big15}

In this section, we evaluate our approach on an improved research benchmark introduced in~\cite{li2025revisiting}, which features controlled and predefined drift occurrences. Although this type of drift is relatively simple and does not capture the complexity of real-world drift observed in MB-24+, we include this evaluation to enable direct comparison with the results reported in~\cite{li2025revisiting}. The following subsections first describe the datasets and the corresponding source-target configurations, followed by the performance results for each step of our framework.
\subsection{Dataset and Experiment Setup}

\subsubsection{Datasets} The Microsoft Malware Classification Challenge (BIG-15) dataset~\cite{ronen2018microsoft} is a  benchmark for evaluating ML-based malware detectors. It has  $\text{10,868}$ malware samples labeled across nine malware families. Despite its widespread use, recent findings~\cite{li2025revisiting} suggest that the high degree of similarity among samples from different families may artificially inflate classification performance when evaluations rely on these family labels. To mitigate this issue, the authors in~\cite{li2025revisiting} introduce a new labeling scheme based on graph-level clustering, which produces new clusters with actual drift between them.  In our evaluation, we adopt this improved labeling strategy and use the four distinct clusters assigned for the BIG-15 dataset.

\subsubsection{Source and target datasets preparation} We adopt the same ``leave-one-cluster-out'' evaluation strategy as in prior works. In each experimental run, one cluster is designated as the target domain (representing new malware), while the remaining clusters are treated as the source malware. This process is repeated using Cluster 0, Cluster 1, and Cluster 2 as the target to ensure the generalizability of our findings.

We follow the same setup as in~\cite{li2025revisiting} to construct the datasets, reusing the benign Windows PE files from the MB-24+ experiments. Among the $16{,}000$ benign samples, $8{,}000$ are randomly assigned to the source domain and the remaining $8{,}000$ to the target domain. Because the BIG-15 dataset lacks temporal information, conventional random sampling is used to divide the data. Specifically, the source domain is split into a training set ($75\%$) and a testing set ($25\%$), while the target domain is split into training and testing subsets ($50\%$ each). To mitigate spatial bias~\cite{pendlebury2019tesseract}, the malware-to-benign ratio is preserved across all partitions within each domain. For the image-based representation, the source and target domains exhibit average ratios of approximately $0.9{:}1$ and $0.38{:}1$, respectively. For the CFG-based representation, the corresponding average ratios are about $1{:}1$ for the source domain and $0.6{:}1$ for the target domain.

\subsection{Evaluation Results}

\subsubsection{Evaluation of Step I}

We evaluate the performance of Step~I along with the same baseline methods used in the previous experiments. Detailed implementation settings for all approaches are provided in Appendix~\ref{app_big15}. Each model is trained on the labeled source training set and the unlabeled target training set, and evaluated on the corresponding target testing set. Performance across the three target clusters is summarized in Table~\ref{big-15-step1}. Step~I of LFreeDA consistently outperforms all four baselines or achieves comparable results to the best-performing baseline when averaged across the three tasks. While AdvDA and DAN (MMD) perform competitively, their success is attributed to the relatively simple benchmark scenario involving only a single novel cluster in the target domain. 

\begin{table}[t]
\caption{Average classification accuracy and F1 after Step I of LFreeDA and baselines on the BIG-15 dataset. Columns correspond to three adaptation tasks with different target malware clusters.}
\label{big-15-step1}
\resizebox{\linewidth}{!}{
\begin{tabular}{@{}crrrrrrrr@{}}
\toprule
\multirow{2}{*}{Methods} & \multicolumn{2}{c}{Cluster 0}                         & \multicolumn{2}{c}{Cluster 1}                         & \multicolumn{2}{c}{Cluster 2}                         & \multicolumn{2}{c}{Avg}                               \\ \cmidrule(l){2-9} 
                         & \multicolumn{1}{c}{Accuracy} & \multicolumn{1}{c}{F1} & \multicolumn{1}{c}{Accuracy} & \multicolumn{1}{c}{F1} & \multicolumn{1}{c}{Accuracy} & \multicolumn{1}{c}{F1} & \multicolumn{1}{c}{Accuracy} & \multicolumn{1}{c}{F1} \\ \midrule
AdvDA + GIN              & 92.3                         & 92.2                   & 90.6                         & 90.6                   & 90.6                         & \textbf{90.0}          & 91.2                         & 90.9                   \\
AdvDA + CNN              & 91.8                         & 91.6                   & 90.4                         & 88.5                   & 87.9                         & 81.4                   & 90                           & 87.2                   \\
DAN + GIN                & 93.4                         & 93.3                   & 91.2                         & \textbf{91.2}          & 91.0                         & 89.3                   & 91.9                         & \textbf{91.3}          \\
DAN + CNN                & 91.7                         & 91.7                   & 90.2                         & 88.3                   & 91.5                         & 85.9                   & 91.1                         & 88.6                   \\
LFreeDA: Step I          & \textbf{94.4}                & \textbf{94.2}          & \textbf{92.5}                & 90.8                   & \textbf{92.5}                & 88.3                   & \textbf{93.1}                & 91.1                   \\ \bottomrule
\end{tabular}
}
\end{table}

\begin{figure}[t]
  \centering
  \includegraphics[width=0.9\linewidth]{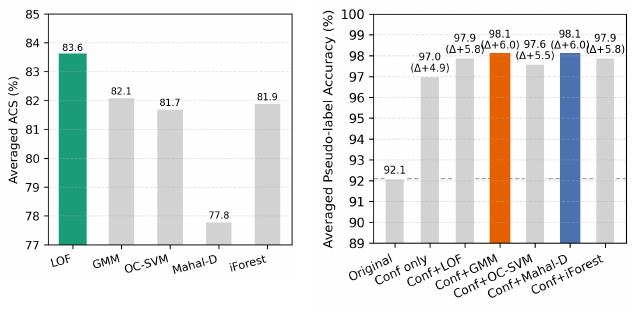}
  \caption{ 
The left graph shows the averaged ACS scores for each outlier detection method combined with confidence filtering on BIG-15. The right graph
compares averaged pseudo-label accuracy: original, with confidence filtering, with
confidence filtering + outlier detection.  }
  \label{big15_step2}
\end{figure}

\begin{figure}[t]
  \centering
  \includegraphics[width=0.9\linewidth]{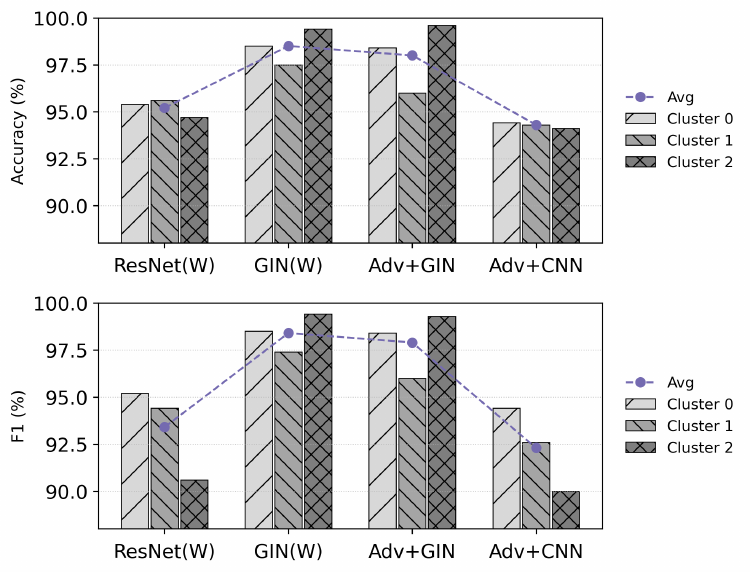}
  \caption{ Average classification accuracy (top) and F1 (bottom) obtained after Step III of LFreeDA on the BIG-15 dataset. ResNet (W): warm-start ResNet-50; GIN (W): warm-start GIN. }
\label{big-15-step3}
\end{figure}

\subsubsection{Evaluation of Step II}  Step II on the BIG-15 benchmark follows the same pseudo-label selection setup as in the MB-24+ experiments (confidence threshold = $0.95$, top $20\%$ outliers across five detection algorithms). Figure \ref{big15_step2} summarizes the averaged results over the three adaptation tasks: the left panel shows the Accuracy-Coverage Score (ACS) for each outlier-detection method, and the right panel reports the corresponding average pseudo-label accuracy. Detailed per-task accuracy and coverage statistics are provided in Appendix Table \ref{app:big15_outlier_stats}.

LOF again achieves the highest ACS on BIG-15, retaining $69\%$ of samples with a labeling accuracy of $97.9\%$, a $5.8\%$ improvement over the initial pseudo-label accuracy of $92.1\%$.

\subsubsection{Evaluation of Step III (final results)}For the BIG-15 benchmark, Step~III trains each adaptation model described in Section~\ref{training} using the labeled source training set together with the pseudo-labeled target training set obtained from Step~II. The adapted models are then evaluated on the corresponding target testing set. Implementation details are provided in Appendix~\ref{app_big15}. The classification accuracy and macro-F1 scores for the three target tasks are reported in Figure~\ref{big-15-step3}. Our observations are as follows:
\begin{itemize}
    \item Models trained on CFG representations consistently outperform those trained on image representations.
    \item Warm-start training achieves accuracy comparable to the AdvDA method, likely due to the large volume of selected pseudo-labeled samples (approximately $\text{4,423}$ on average).
    \item The highest average accuracy achieved is $98\%$, comparable to the results reported in~\cite{li2025revisiting}, where AdvDA is trained using $50$ true labels from the target domain.
\end{itemize}

\section{Evaluation: Controlled Drift on Family Classification }\label{evaluation:family}
We evaluated our framework on malware family classification, where source and target domains share the same families. Our results show that the size of unlabeled target data affects performance and reveal the minimum samples per class needed for effective adaptation.

\subsection{Datasets and Experiment Setup}

\subsubsection{Datasets}  We use the MalwareDrift dataset~\cite{ma2021comprehensive, wadkar2020detecting}, which includes pre- and post-drift samples from seven malware families spanning trojans, worms, adware, and backdoors. Prior work~\cite{ma2021comprehensive} shows that models trained on pre-drift data perform poorly on post-drift samples, achieving only {$42\%$} accuracy.

\subsubsection{Source and target dataset preparation} 
The source domain includes pre-drift samples from seven malware families and benign Windows files, while the target domain contains post-drift samples from the same families and different benign files. For each malware family, the post-drift data is time-ordered and split in half: the first half is used for target training and the second half for target testing. These per-family splits are then combined to form the final target training and testing sets, preserving class distribution of the original post-drift segment and avoiding spatial bias~\cite{pendlebury2019tesseract}. All pre-drift data is used for source training. Temporal bias is avoided by ensuring that all training data precedes testing data~\cite{pendlebury2019tesseract}. Sample counts per class are detailed in Table~\ref{md}.

\begin{table}[t]
\caption{Number of samples per class in the source and target datasets}
\label{md}
\resizebox{\linewidth}{!}{
\begin{tabular}{@{}lrrrrrrrr@{}}
\toprule
\textbf{Subsets} & \multicolumn{1}{c}{\textbf{benign}} & \multicolumn{1}{c}{\textbf{Bifrose}} & \multicolumn{1}{c}{\textbf{Ceeinject}} & \multicolumn{1}{c}{\textbf{Obfuscator}} & \multicolumn{1}{c}{\textbf{Vbinject}} & \multicolumn{1}{c}{\textbf{Vobfus}} & \multicolumn{1}{l}{\textbf{Winwebsec}} & \multicolumn{1}{l}{\textbf{Zegost}} \\ \midrule
Source training  & 500                                 & 171                                  & 90                                      & 143                                     & 379                                   & 64                                  & 218                                    & 180                                 \\
Target training  & 250                                 & 53                                   & 229                                     & 30                                      & 327                                   & 109                                 & 134                                    & 57                                  \\
Target testing   & 250                                 & 53                                   & 229                                     & 30                                      & 327                                   & 109                                 & 134                                    & 57                                  \\ \bottomrule
\end{tabular}}
\end{table}

\begin{table}[t]
\centering
\caption{Accuracy and F1  obtained
after Step I and Step III of LFreeDA on the MalwareDrift dataset.}\label{md_step1_3}

\begin{minipage}[t]{0.48\linewidth}
\centering
(a) Results after Step I\\[2pt]
\resizebox{\linewidth}{!}{
\begin{tabular}{@{}crr@{}}
\toprule
\multirow{2}{*}{Methods} & \multicolumn{2}{c}{Target testing} \\ \cmidrule(l){2-3} 
                         & Acc              & F1              \\ \midrule
AdvDA + GIN              & 45.9             & 44.6            \\
AdvDA + CNN              & 45.2             & 43.1            \\
DAN + GIN                & 45.7             & 44.2            \\
DAN + CNN                & 44.9             & 43.9            \\
LFreeDA: Step I          & \textbf{47.2}    & \textbf{46.1}   \\ \bottomrule
\end{tabular}}
\end{minipage}\hfill
\begin{minipage}[t]{0.48\linewidth}
\centering
(b) Results after Step III\\[2pt]
\resizebox{\linewidth}{!}{
\begin{tabular}{@{}crr@{}}
\toprule
\multirow{2}{*}{Methods} & \multicolumn{2}{c}{Target testing} \\ \cmidrule(l){2-3} 
                         & Acc              & F1              \\ \midrule
Warm-start ResNet-50     & 46.4             & 45.9            \\
Warm-start GIN           & 52.5             & 46.0            \\
AdvDA + GIN              & \textbf{52.9}    & \textbf{50.5}   \\
AdvDA + CNN              & 47.6             & 45.2            \\ \bottomrule
\end{tabular}}
\end{minipage}

\end{table}

\subsection{Evaluation Results}

\subsubsection{Evaluation of Step I} We adopt the same baseline methods described in Section~\ref{eva:big15}, with implementation details provided in Appendix~\ref{app_MD_imple}. Table~\ref{md_step1_3} (a) summarizes the results. Our Step I outperforms all four baselines. However, all methods show reduced performance due to the limited number of target training samples per class. Three classes have fewer than 60 target samples each.

\subsubsection{Evaluation of Step II} In contrast, pseudo-label selection methods show notable improvement, achieving $69\%$ accuracy on pseudo-labels selected with confidence filtering, a $20\%$ gain over the original accuracy. Among the methods, LOF has the highest ACS score of $56.4\%$ and further improves accuracy to $77\%$. Detailed accuracy and ACS scores are presented in Figure~\ref{md_step2} in the Appendix.

\subsubsection{Evaluation of Step III (final results)} Classification accuracy and F1 are reported in Table~\ref{md_step1_3} (b), and implementation details in Appendix~\ref{app_MD_imple}. CFG-based representation outperforms in both the warm-start and AdvDA methods. Notably, AdvDA + GIN achieves the best result, improving accuracy by $5.7\%$ over Step I alone.

\subsection{Further analysis} 
\begin{figure}[t]
  \centering
  \includegraphics[width=0.85\linewidth]{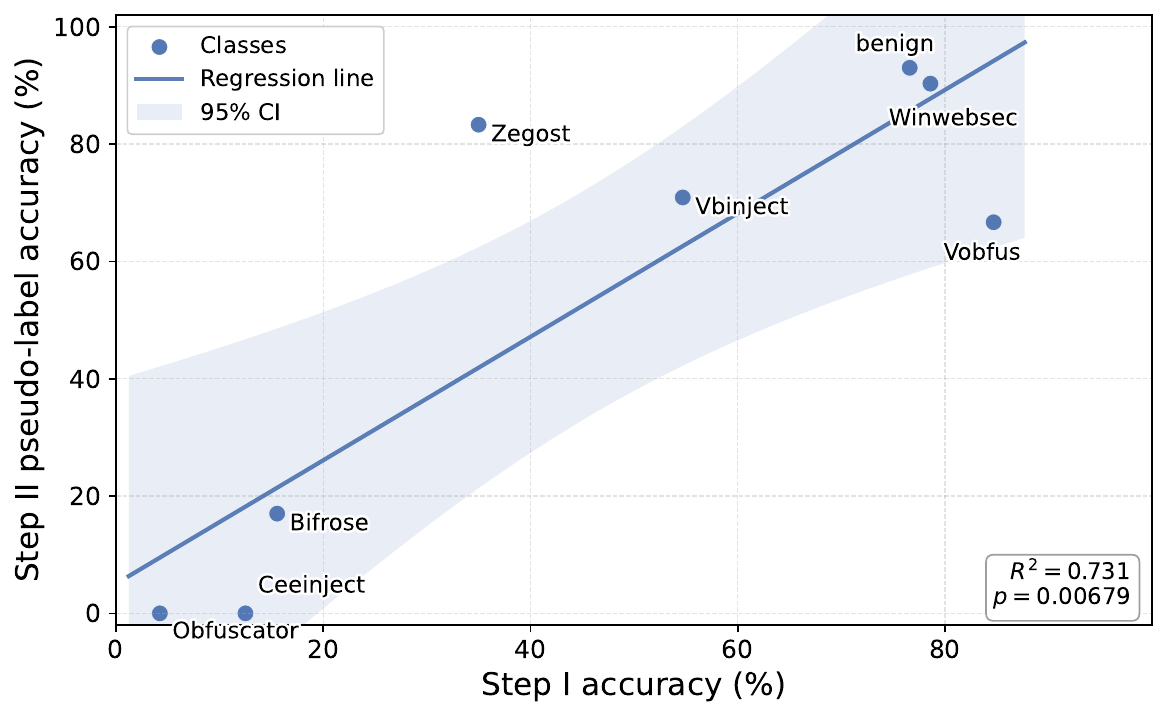}
  \caption{Correlation between initial Step I accuracy and Step II pseudo-label accuracy, grouped by predicted class. $R^2 \in [0,1]$ with values closer to $1$ indicating a stronger linear relationship; $p \le 0.05$ indicates a statistically significant correlation.}\label{correction}
\end{figure}

While our framework achieves $52.9\%$ accuracy without any post-drift labels, it underperforms compared to AdvDA trained with 10 labeled samples per class in the post-drift set~\cite{li2025revisiting}. To investigate this gap,  we analyze pseudo-label quality.  Although the overall accuracy of selected pseudo-labels is $77\%$, class-wise accuracy varies significantly. We observe a strong correlation between Step I’s prediction accuracy and the pseudo-label accuracy after Step II across all predicted classes, as shown in Figure \ref{correction}, where each point represents a predicted malware family. This suggests that inaccurate predictions in Step I lead to noisy pseudo-labels.

\begin{figure}[t]
  \centering
  \includegraphics[width=0.85\linewidth]{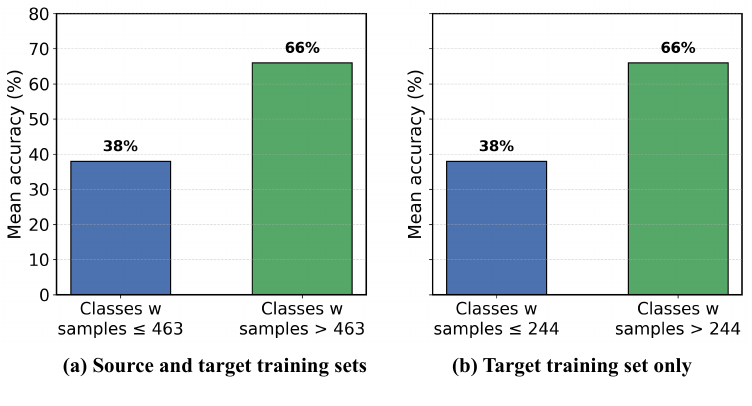}
  \caption{Impact of the low-sample classes and high-sample classes on Step I's performance.}\label{size}
\end{figure}

We hypothesize that this limitation stems from data scarcity in certain classes, which is not considered in the original MaxDIRep work~\cite{li2023maximal}. 
To test this, we rank classes by sample count in the target training set and divide them into ``low'' ($<75^{\text{th}}$ percentile) and ``high'' ($\geq75^{\text{th}}$ percentile) groups. Figure~\ref{size} (b) compares the mean accuracy of predicted classes within these groups. Figure~\ref{size} (a) presents the same analysis using combined source and target training data.  Results show a clear accuracy boost for classes with $\geq244$ target samples or $\geq463$ combined samples, supporting our hypothesis that larger unlabeled datasets improve  Step I performance.

\section{Practical Insights}\label{evaluation:practical}

\textbf{Insight 1:} \textit{Our framework's backbone is Step I, which demonstrates strong performance when each class has sufficient unlabeled data.}

In our MalwareDrift experiments, we observed that Step I consistently achieved higher accuracy in classes with more data, while performance degraded for classes with fewer samples. This difference in Step I accuracy, in turn, translated into differences in pseudo-label quality in Step II: classes that were predicted more accurately in Step I also produced higher-quality pseudo-labels after selection. Empirically, classes with at least $244$ target samples and $463$ total (source + target) samples exhibited a noticeable accuracy boost. While not definitive cutoffs, these values indicate the approximate data scale required for reliable pseudo-label generation.

Estimating the amount of data per class is feasible for an unlabeled target domain. Since the source and target domains share the same label space, the number of classes is known a priori. Clustering algorithms such as K-means can be applied to the image representations of the target training data to approximate the number of samples per class. The resulting cluster sizes provide a reasonable estimate of class-wise data availability. 

For underrepresented classes, synthetic data generation presents a promising solution. Since our Step I uses image-based representations of malware, generative models for image data like autoencoders~\cite{wan2017variational} and GANs~\cite{goodfellow2016deep} can be used to augment these classes. Although incorporating synthetic data is beyond the scope of this work, it remains a valuable direction for future research.

\textbf{Insight 2:} \textit{Local outlier factor achieves the highest ACS score across datasets.}

In practical deployments, the lack of ground-truth labels in target training data makes it impossible to directly assess the performance of outlier detection algorithms. To address this, we evaluated five outlier detection methods across three malware datasets, using the target training labels only to measure pseudo-label accuracy. We further introduced the ACS score, which quantifies an algorithm’s ability to select a large portion of samples while maintaining high accuracy. Among the five methods, LOF consistently achieved the highest ACS scores, making it the most suitable choice for use in Step II.

\textbf{Insight 3:} \textit{AdvDA + GIN overall is the best DA method in Step III. }

AdvDA combined with GIN achieves the best or near-best performance across all evaluated datasets. This suggests that practitioners can confidently use the CFG-based representation with AdvDA + GIN as the default adaptation method in Step III. In this setting, the model is trained on CFG representations from both the source and selected target training data, using ground-truth labels for the source and pseudo-labels from Step II for the selected target samples. We also find no clear advantage between the warm-start ResNet50 and AdvDA + CNN for the image-based representation: their performance is comparable on both MB-24+ and MalwareDrift, with warm-start ResNet50 showing a slight edge on the BIG-15 dataset.

Based on our insights and experimental analysis, we articulate some \textbf{deployment recommendations}.

Before deploying LFreeDA, practitioners should estimate class-wise data distribution in the unlabeled target training set using the clustering approach described above. LFreeDA performs best when each class has at least $244$ samples; otherwise, additional unlabeled or synthetic data is needed. After Step~I, we suggest confidence-based filtering combined with LOF to select high-quality pseudo-labels, as this offers the best balance between label quality and coverage. In Step~III, adaptation should use AdvDA + GIN with the CFG-based representation, which achieves the highest performance in our experiments. These recommendations are derived from evaluations on three datasets covering diverse sizes and class distributions. The thresholds of $244$ target samples and $463$ combined samples were calibrated on MalwareDrift, a dataset with severe class imbalance and data scarcity (Table~\ref{md}). Set under such adverse conditions, these thresholds are conservative and should generalize to larger, balanced datasets.


\section{Discussion}
While this work does not focus on adversarial attacks against malware classifiers, we outline strategies to improve LFreeDA’s robustness. Adversarial training can be incorporated in Step III by generating adversarial malware examples from the original binaries in both domains and including them during training. Because the target training set is initially unlabeled, adversarial examples in Step I can only be generated from the source training set. Future work will explore integrating this defense into LFreeDA.

Step II is method-agnostic, and recent methods such as adaptive confidence thresholding with mixup regularization~\cite{alam2025adapt} could further improve pseudo-label selection. Mixup regularization combines the features and labels of two samples to enhance model generalization and calibration. Nevertheless, our approach is efficient and achieves strong performance, as demonstrated in this study.

LFreeDA assumes that the source and target domains share the same label space. Extending it to open-set scenarios, where the two domains differ in their label spaces, is challenging. Our evaluation shows that LFreeDA performs well when each class has sufficient samples (at least 244 empirically), which requires estimating the number of samples per class in the target training set. In the closed-set setting, where the number of classes is known, class-wise sample sizes can be estimated more accurately via clustering. In contrast, in open-set settings, the number of clusters is unknown and these sizes are harder to infer, as unknown classes disrupt cluster structure~\cite{bendale2016towards}. Therefore, we limit our study to the closed-set problem.

If data scarcity is not an issue for known target classes, another direction is to strengthen Step I for the open-set setting. In such settings, state-of-the-art UDA methods~\cite{panareda2017open, saito2018open} can (1) accurately classify target samples from known classes and (2) detect and label samples from unknown classes as ``unknown.'' To move LFreeDA toward the second capability, Step I could be augmented with contrastive losses on source labels to make DIReps form tight, well-separated clusters for known families, and one-class models (e.g., one-class SVMs) could be trained on source DIReps for each family. At test time, target samples whose DIReps lie within the support of a single known-class model would be classified as that family, while samples falling outside the support of all known-class models would be assigned an ``unknown'' label.


\section{Related Work}

\textbf{Drift mitigation with pseudo-labels.} DroidEvolver~\cite{xu2019droidevolver} uses an ensemble of online learning algorithms, initially trained on labeled source data and later retrained with generated pseudo-labels. Predictions are obtained by aggregating decision scores from models not flagged as ``aging'', where aging is detected via a juvenilization indication (JI) score; DroidEvolver++~\cite{kan2021investigating} improves upon this framework by refining the ensemble decision function and pseudo-label integration. However, as noted in~\cite{kan2021investigating}, self-learning based on a model’s own predictions can lead to self-poisoning, where errors propagate rapidly under severe concept drift. LFreeDA differs in two key aspects: (1) it uses UDA to generate pseudo-labels instead of online learning algorithms, which are prone to forgetting previously acquired knowledge~\cite{kirkpatrick2017overcoming,kemker2018measuring} and limited in capturing shared features between pre-drift and post-drift data; and (2) it avoids iterative self-training, which risks self-poisoning from inaccurate labels. Instead, we periodically train a new Step~I model from scratch on both source and newly collected target data. ADAPT~\cite{alam2025adapt} is a more recent self-training approach that mitigates self-poisoning through adaptive confidence thresholding and mixup-based confidence calibration. ADAPT and LFreeDA differ in two fundamental ways. ADAPT uses target data only for inference to generate pseudo-labels, not to train the model that produces them, while LFreeDA incorporates unlabeled target data via UDA.  ADAPT trains its final classifier by warm-starting on tabular features, while LFreeDA uses the more advanced AdvDA method on CFG-based representations, which has been shown to outperform warm-start training on tabular malware features~\cite{li2025revisiting}.

\textbf{Human/Sandbox-assisted drift mitigation}. Most of the work belongs to this category, where a small budget of human labeling is allowed. Typically, a drift detection algorithm~\cite{jordaney2017transcend, yang2021cade, tripathi2025towards, botacin2025towards}  or a sample selection algorithm~\cite{chen2023continuous} is designed to identify drifted samples, which are then forwarded for human annotation/sandbox analysis and subsequently incorporated into model retraining pipelines. While these approaches reduce annotation effort compared to full labeling, they still require an additional drift-detection step and manual labeling/sandbox analysis. In contrast, LFreeDA removes both requirements entirely, allowing the model to adapt to drift automatically. 


\textbf{Unsupervised DA.} Two approaches closely related to MaxDIRep are DSAN~\cite{stojanov2021domain} and DSN~\cite{bousmalis2016domain}. DSAN incorporates domain-specific information alongside the input data to learn the DIRep. In contrast, MaxDIRep learns the DIRep without requiring any domain-specific input. The key distinction between DSN and MaxDIRep is the constraints applied to learn DIRep and DDRep.  DSN employs an orthogonality constraint between those representations to ensure they are distinct.  MaxDIRep imposes a stronger constraint aimed at minimizing the information content of the DDRep, which leads to superior performance compared to both DSN and DSAN~\cite{li2023maximal}. Other related approaches incorporate pseudo-labels with UDA~\cite{chen2020adversarial, zou2018unsupervised}, although these are confined to the computer vision domain.

\section{Conclusion}

In this paper, we introduced LFreeDA, a framework for adapting malware detection models to concept drift without relying on labeled drifted samples, thereby eliminating the need for manual annotation. Leveraging scalable unsupervised domain adaptation on malware image representations, LFreeDA generates pseudo-labels and retains only high-quality predictions while balancing the quantity of selected samples. The selected pseudo-labeled data are then used to train the final malware classifier with DA methods, where we also exploit the advantages of CFG-based representations to improve performance. Extensive evaluations on benchmark and real-world datasets show that LFreeDA matches state-of-the-art methods that require up to $300$ labeled drifted samples, given sufficient unlabeled data per class. Our empirical findings also provide practical guidance for real-world deployment.

\noindent\textbf{Acknowledgements}. The work has been supported by the National Science Foundation (NSF) under Grants 2229876 and 2112471.

\bibliographystyle{IEEEtranS}
\bibliography{main}

\appendices
\section{Image Feature Preprocessing Details}\label{app_image}
Following~\cite{nataraj2011malware}, the image width is determined based on file size (e.g., $32, 64, 128, 256, 384, 512, 768, 1024$), and the height is computed as $\mathit{\frac{Total\ bytes}{Width}}$. Bytes are laid out left-to-right, top-to-bottom, so consecutive bytes map to adjacent pixels, preserving local byte structure; any incomplete row is implicitly zero-padded. To ensure uniform input dimensions, all images are resized to $56\times56$ using LANCZOS interpolation, which preserves local structures better than cropping or padding. This configuration is kept consistent with~\cite{li2025revisiting}.

\section{Outlier Detection Algorithms Integration}\label{app_ODA}

\textbf{Local outlier factor (LOF)}. LOF~\cite{Breunig2000} is a density-based method that detects outliers through nearest neighbor analysis. We use LOF to assess the local density of each sample relative to its $k$ nearest neighbors. The density is computed from the distances to these neighbors. By comparing a sample’s density to that of its neighbors, LOF distinguishes between regions of similar density and points that exhibit significantly lower density, which are flagged as outliers.

\textbf{Gaussian mixture model (GMM)}. GMM provides a probabilistic description of the data density and therefore can be repurposed for outlier detection~\cite{Bishop2006}.  We follow the same approach for use in our pseudo-label selection module. After fitting the GMM model with all the input data,  each data point receives a log-likelihood score. Because the log-likelihood is monotonically related to the posterior inlier probability, data points that fall in the lower tail of the likelihood distribution can be interpreted as statistical outliers under the fitted model. We thus flag as outliers the samples whose likelihood falls below a specified quantile (e.g., the lowest $20\%$).

\textbf{One-class SVM (OC-SVM)}.
OC-SVM~\cite{Scholkopf2001} learns a decision function that captures the regions in the feature space where the majority of the training data lie, effectively modeling the ``normal'' class. The algorithm maps input data into a high-dimensional feature space using a kernel function and attempts to separate the origin from the data points with maximum margin. For use in our pseudo-label selection module, we directly fit the OC-SVM with input data; then data points predicted with  $-1$ are flagged as outliers.

\textbf{Mahalanobis-distance detector (Mahal-D)}.
The Mahalanobis distance is a statistical measure used to detect outliers by evaluating how far a point lies from the mean of a distribution, while accounting for correlations between features~\cite{RousseeuwHubert2018}. For use in our pseudo-label selection module, we first compute the mean and covariance of the entire input and then compute the Mahalanobis distance from each data point to the mean.  To determine outliers, we compare these distances against a threshold derived from the $\chi^2$ distribution with a number of degrees of freedom equal to the number of features. Observations with distances exceeding this threshold are flagged as outliers.

\textbf{Isolation forest (iForest).} iForest~\cite{Liu2008IsolationForest} is an ensemble-based anomaly detection method. The algorithm constructs multiple binary trees (isolation trees) by recursively selecting random features and split values. The average path length (from the root node to the terminating node) to a given data point across all trees serves as an outlier score. Shorter paths indicate a higher likelihood of being an outlier since random partitioning produces noticeably shorter paths for outliers. We directly fit the model with input data; then, data points predicted with  $-1$ are flagged as outliers.

\section{Additional Details for MB-24+ Evaluation}

\subsection{Implementation Details of Unsupervised DA Methods}\label{app_mb_DA}
Step I of LFreeDA: The generator follows the architecture: (\texttt{Conv2D}, \verb|MaxPooling|, \verb|Conv2D|, \verb|MaxPooling|). The two \verb|Conv2D| layers use $3 \times 3$ filters with $32$ and $64$ channels, respectively. The output of the generator is the DIRep with shape $12 \times 12 \times 64$. The classifier consists of a \verb|Flatten| layer followed by two fully connected layers: \verb|FC_1| with $256$ neurons and \verb|FC_OUT| for label prediction. The discriminator includes a \verb|Flatten| layer, two dense layers with $1024$ hidden units each, and a final \verb|Softmax| layer for domain prediction.
The encoder consists of two \verb|Conv2D| layers with stride $2$, each using $3 \times 3$ filters with $32$ and $64$ channels, respectively. This is followed by two \verb|Conv2D| layers for computing $z\_mean$ and $z\_log\_var$, each with $2$ filters of size $3 \times 3$. A \verb|Sampling| layer then produces the latent representation $z$ (DDRep) with shape $12 \times 12 \times 2$. The decoder concatenates DDRep with DIRep, forming an input of shape $12 \times 12 \times 66$. Its architecture is: (\verb|Conv2D|, \verb|Conv2DTranspose| $\times 4$). The initial \verb|Conv2D| uses $64$ filters of size $3 \times 3$. The next three \verb|Conv2DTranspose| layers use $3 \times 3$ filters with $64$, $64$, and $32$ channels, respectively, progressively upsampling the feature map. A final \verb|Conv2DTranspose| layer with $3$ filters of size $3 \times 3$ and sigmoid activation reconstructs the image. The model is trained using the Adam optimizer with a learning rate of $\mathit{1e{-}3}$ for $60$ epochs and a batch size of $32$. The loss coefficients are set as follows: $\mathcal{L}_g: 0.1$, $\mathcal{L}_{\text{kl}}: \frac{1}{20000}$, and $\mathcal{L}_{\text{recon}}: 0.05$.

We use the open-source implementations of AdvDA + GIN, AdvDA + CNN, DAN + GIN, and DAN + CNN provided in the artifact of~\cite{li2025revisiting}.  We modify the computation of the classification loss $\mathcal{L}_c$ to ensure it is applied only to the source domain. Below, we summarize the implementation details of each model.

AdvDA + GIN: The generator consists of three \verb|GIN| layers, followed by a global average pooling layer and a dense layer with $256$ neurons. The classifier includes two fully connected layers: \verb|FC_1| with $256$ neurons and \verb|FC_OUT| for label prediction. The discriminator comprises two dense layers with $256$ hidden units each, followed by a \verb|Softmax| layer for domain prediction. The model is trained using the Adam optimizer with a learning rate of $\mathit{1e{-}3}$ for $60$ epochs and a batch size of $16$. The loss coefficient $\mathcal{L}_g$ is set to $0.1$. 

AdvDA + CNN: The generator follows the architecture: \verb|Conv2D|, \verb|MaxPooling|, \verb|Conv2D|, \verb|MaxPooling|, \verb|Flatten|. The two \verb|Conv2D| layers use $3 \times 3$ filters with $32$ and $64$ channels, respectively. The classifier consists of two fully connected layers: \verb|FC_1| with $256$ neurons and \verb|FC_OUT| for label prediction. The discriminator includes two dense layers with $1024$ hidden units each, followed by a \verb|Softmax| layer for domain prediction. The model is trained using the Adam optimizer with a learning rate of $\mathit{1e{-}3}$ for $60$ epochs and a batch size of $32$. The loss coefficient $\mathcal{L}_g$ is set to $0.1$. 

DAN + GIN: The generator and classifier architectures are identical to those used in AdvDA + GIN. The model is trained using the Adam optimizer with a learning rate of $\mathit{1e{-}3}$ for $60$ epochs and a batch size of $16$. The domain adaptation is guided by the Maximum Mean Discrepancy (MMD) loss, with a coefficient set to $1$, following~\cite{long2015learning}. The MMD loss is computed on the hidden graph representations produced by the generator. As in~\cite{li2025revisiting}, we use the RBF kernel for MMD computation. 

DAN + CNN: The generator and classifier architectures are identical to those used in AdvDA + CNN. The model is trained using the Adam optimizer with a learning rate of $\mathit{1e{-}3}$ for $60$ epochs and a batch size of $32$. The coefficient of the MMD loss is set to $1$.

\subsection{Implementation Details of Outlier Detection Algorithms}\label{app_mb_out}

Note that our outlier detection algorithm operates on the DIReps of samples belonging to the same predicted class. After reshaping, each DIRep has a dimensionality of $12 \times 12 \times 64 = 9216$ features. Due to the challenges associated with high-dimensional data, we apply Principal Component Analysis (PCA) to reduce the feature dimension from $9216$ to $50$ before feeding the data into the outlier detection algorithm. All outlier detection models are trained on the PCA-reduced features, with the outlier fraction set to $0.2$.

\subsection{Implementation Details of DA Methods in Step III}\label{app_mb_train}
Warm-start ResNet-50: To adapt a pretrained ResNet-50 with ImageNet for binary classification (malware vs. benign), we modify the original architecture by removing its final classification layer (originally with $1000$ neurons). We then add a \verb|GlobalAvgPool| layer, followed by a fully connected layer with $256$ neurons, and a final output layer with $2$ neurons. The model is trained using the Adam optimizer with a learning rate of $\mathit{1e{-}3}$ for $60$ epochs.

Warm-start GIN: The GIN backbone consists of three stacked \verb|GINConv| layers ($\varepsilon = 0$), each with a two-layer \verb|MLP| $[128, 128]$. 
Node embeddings are aggregated using \verb|GlobalAvgPool|, followed by a dense layer of 128 units with \verb|ReLU| activation, 
a \verb|Dropout| layer, and a final \verb|Softmax| output layer with two neurons. The model is first pretrained on the source domain and then fine-tuned on high-confidence pseudo-labeled target samples, with the first layer frozen following the setup in~\cite{li2025revisiting}.

AdvDA + GIN and AdvDA + CNN: We use the publicly available implementations provided in~\cite{li2025revisiting}. 

Lower and upper bounds: Both bounds are implemented using the same base architectures described above: ResNet-50 for image-based representation and GIN for CFG-based representation.

\subsection{Additional Results}

Table~\ref{app:mb_outlier_stats} reports the per-update statistics for pseudo-label accuracy, coverage, and ACS for each outlier detection method combined with confidence-based filtering.

Table~\ref{mb24_ratio} summarizes the malware-to-benign ratios in the source, target training, and target testing partitions for each post-drift testing month.

\begin{figure*}[t]
  \centering
  \includegraphics[width=0.9\linewidth]{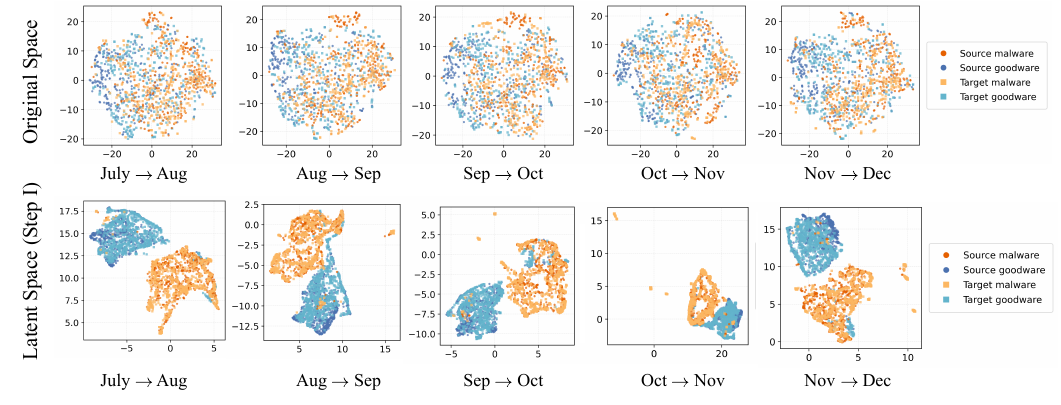}
  \caption{T-SNE visualization for the original space, and the latent space of LFreeDA Step I (DIRep features) for all the model adaptation tasks on MB-24+.}
  \label{app_mb24_tsne_full}
\end{figure*}

Figure~\ref{app_mb24_tsne_full} shows the t-SNE visualizations of the original feature space and the Step~I latent space for all adaptation tasks.

\begin{table*}[t]
\caption{Detailed per-update performance of each outlier-detection method used with confidence-based filtering on the MB-24+ dataset, including pseudo-label accuracy, coverage, and Accuracy-Coverage Score (ACS).}
\label{app:mb_outlier_stats}
\resizebox{\linewidth}{!}{
\begin{tabular}{@{}crrrrrrrrrrrrrrr@{}}
\toprule
\multirow{2}{*}{Methods} & \multicolumn{3}{c}{Aug Testing}                                                       & \multicolumn{3}{c}{Sep Testing}                                                       & \multicolumn{3}{c}{Oct Testing}                                                       & \multicolumn{3}{c}{Nov Testing}                                                       & \multicolumn{3}{c}{Dec Testing}                                                       \\ \cmidrule(l){2-16} 
                         & \multicolumn{1}{c}{Accuracy} & \multicolumn{1}{c}{Coverage} & \multicolumn{1}{c}{ACS} & \multicolumn{1}{c}{Accuracy} & \multicolumn{1}{c}{Coverage} & \multicolumn{1}{c}{ACS} & \multicolumn{1}{c}{Accuracy} & \multicolumn{1}{c}{Coverage} & \multicolumn{1}{c}{ACS} & \multicolumn{1}{c}{Accuracy} & \multicolumn{1}{c}{Coverage} & \multicolumn{1}{c}{ACS} & \multicolumn{1}{c}{Accuracy} & \multicolumn{1}{c}{Coverage} & \multicolumn{1}{c}{ACS} \\ \midrule
LOF                      & 89.07                        & 61.9                         & \textbf{75.5}           & 86.5                         & 63.6                         & \textbf{75.1}           & 86.1                         & 60.4                         & \textbf{73.3}           & 90.1                         & 63.7                         & \textbf{76.9}           & 89.2                         & 59.9                         & \textbf{74.6}           \\
GMM                      & 88.8                         & 59.2                         & 74                      & 86.5                         & 62.1                         & 74.3                    & 86.6                         & 57.8                         & 72.2                    & 90.6                         & 62.0                         & 76.3                    & 90.4                         & 55.8                         & 73.1                    \\
OC-SVM                   & 88.8                         & 59.1                         & 74                      & 86.1                         & 62.0                         & 74.1                    & 86.4                         & 57.6                         & 72                      & 90.5                         & 61.9                         & 76.2                    & 90.3                         & 55.4                         & 72.9                    \\
Mahal-D                  & 88.9                         & 57.1                         & 73                      & 86.5                         & 58.5                         & 72.5                    & 87.0                         & 51.3                         & 69.2                    & 90.8                         & 57.9                         & 74.4                    & 91.3                         & 50.2                         & 70.8                    \\
iForest                  & 88.4                         & 59.1                         & 73.8                    & 86.2                         & 61.9                         & 74.1                    & 86.7                         & 57.5                         & 72.1                    & 90.4                         & 61.7                         & 76.1                    & 90.1                         & 55.5                         & 72.8                    \\ \bottomrule
\end{tabular}
}
\end{table*}

\begin{table}[h]
\caption{Malware-to-benign ratios in source/target training and testing partitions for each testing month. }
\label{mb24_ratio}
\begin{tabular}{@{}ccrrr@{}}
\toprule
\textbf{Representation} & \textbf{\begin{tabular}[c]{@{}c@{}}Testing \\ Month\end{tabular}} & \multicolumn{1}{c}{\textbf{\begin{tabular}[c]{@{}c@{}}Source\\ Training/Testing\end{tabular}}} & \multicolumn{1}{c}{\textbf{\begin{tabular}[c]{@{}c@{}}Target\\ Training\end{tabular}}} & \multicolumn{1}{c}{\textbf{\begin{tabular}[c]{@{}c@{}}Target \\ Testing\end{tabular}}} \\ \midrule
\multirow{6}{*}{CFG}    & 08/2024                                                           & 0.6:1                                                                                          & 0.5:1                                                                                  & 0.6:1                                                                                  \\
                        & 09/2024                                                           & 0.6:1                                                                                          & 0.6:1                                                                                  & 0.5:1                                                                                  \\
                        & 10/2024                                                           & 0.6:1                                                                                          & 0.5:1                                                                                  & 0.5:1                                                                                  \\
                        & 11/2024                                                           & 0.6:1                                                                                          & 0.5:1                                                                                  & 0.4:1                                                                                  \\
                        & 12/2024                                                           & 0.6:1                                                                                          & 0.4:1                                                                                  & 0.4:1                                                                                  \\
                        & \textbf{Avg}                                                      & \textbf{0.6:1}                                                                                 & \textbf{0.5:1}                                                                         & \textbf{0.5:1}                                                                         \\ \midrule
\multirow{6}{*}{Image}  & 08/2024                                                           & 0.5:1                                                                                          & 0.4:1                                                                                  & 0.4:1                                                                                  \\
                        & 09/2024                                                           & 0.5:1                                                                                          & 0.4:1                                                                                  & 0.3:1                                                                                  \\
                        & 10/2024                                                           & 0.5:1                                                                                          & 0.3:1                                                                                  & 0.3:1                                                                                  \\
                        & 11/2024                                                           & 0.5:1                                                                                          & 0.3:1                                                                                  & 0.3:1                                                                                  \\
                        & 12/2024                                                           & 0.5:1                                                                                          & 0.3:1                                                                                  & 0.3:1                                                                                  \\
                        & \textbf{Avg}                                                      & \textbf{0.5:1}                                                                                 & \textbf{0.3:1}                                                                         & \textbf{0.3:1}                                                                         \\ \midrule
\multicolumn{1}{l}{}    & \multicolumn{1}{l}{}                                              &                                                                                                & \multicolumn{1}{l}{}                                                                   & \multicolumn{1}{l}{}                                                                   \\
\multicolumn{1}{l}{}    & \multicolumn{1}{l}{}                                              &                                                                                                & \multicolumn{1}{l}{}                                                                   & \multicolumn{1}{l}{}                                                                   \\
\multicolumn{1}{l}{}    & \multicolumn{1}{l}{}                                              &                                                                                                & \multicolumn{1}{l}{}                                                                   & \multicolumn{1}{l}{}                                                                   \\
\multicolumn{1}{l}{}    & \multicolumn{1}{l}{}                                              &                                                                                                & \multicolumn{1}{l}{}                                                                   & \multicolumn{1}{l}{}                                                                  
\end{tabular}
\end{table}

\section{Additional Details for BIG-15 Evaluation}

\subsection{Implementation Details}\label{app_big15}
The implementation remains consistent with those outlined in Appendix~\ref{app_mb_DA},~\ref{app_mb_out}, and~\ref{app_mb_train}.

\subsection{Additional Results}

Table~\ref{app:big15_outlier_stats} summarizes the pseudo-label accuracy, coverage, and Accuracy-Coverage Score (ACS) of different outlier-detection methods combined with confidence-based filtering on the BIG-15 adaptation tasks.

\begin{table}[h]
\caption{Detailed per-task performance of each outlier-detection method used with confidence-based filtering on the Big-15 dataset, including pseudo-label accuracy, coverage, and Accuracy-Coverage Score (ACS).}
\label{app:big15_outlier_stats}
\resizebox{\linewidth}{!}{
\begin{tabular}{@{}crrrrrrrrr@{}}
\toprule
\multirow{2}{*}{Methods} & \multicolumn{3}{c}{Cluster 0}                                                                & \multicolumn{3}{c}{Cluster 1}                                                                & \multicolumn{3}{c}{Cluster 2}                                                                \\ \cmidrule(l){2-10} 
                         & \multicolumn{1}{l}{Accuracy} & \multicolumn{1}{c}{Coverage} & \multicolumn{1}{c}{ACS} & \multicolumn{1}{l}{Accuracy} & \multicolumn{1}{c}{Coverage} & \multicolumn{1}{c}{ACS} & \multicolumn{1}{l}{Accuracy} & \multicolumn{1}{c}{Coverage} & \multicolumn{1}{c}{ACS} \\ \midrule
LOF                      & 97.9                         & 72.3                         & \textbf{85.1}           & 97.2                         & 70.8                         & \textbf{84.0}           & 98.7                         & 64.5                         & \textbf{81.6}           \\
GMM                      & 97.7                         & 70.4                         & 84.0                    & 96.9                         & 67.1                         & 82.0                    & 99.6                         & 61.1                         & 80.3                    \\
OC-SVM                   & 97.2                         & 69.9                         & 83.6                    & 96.6                         & 66.8                         & 81.7                    & 98.9                         & 60.6                         & 79.7                    \\
Mahal-D                  & 97.9                         & 61.0                         & 79.4                    & 96.8                         & 57.2                         & 77.0                    & 99.6                         & 54.4                         & 77.0                    \\
iForest                  & 97.5                         & 69.9                         & 83.7                    & 96.5                         & 67.0                         & 81.8                    & 99.6                         & 60.9                         & 80.3                    \\ \bottomrule
\end{tabular}
}
\end{table}

\section{Additional Details for Family Classification}\label{app_MD}

\subsection{Implementation Details}\label{app_MD_imple}

The implementation is the same as described in Appendix~\ref{app_mb_DA},~\ref{app_mb_out}, and~\ref{app_mb_train}, except for modifications to the output layers of the classifiers in Steps I and III, which have been adjusted to support eight classes.

\subsection{Additional Results}
\begin{figure}[h]
  \centering
  \includegraphics[width=0.9\linewidth]{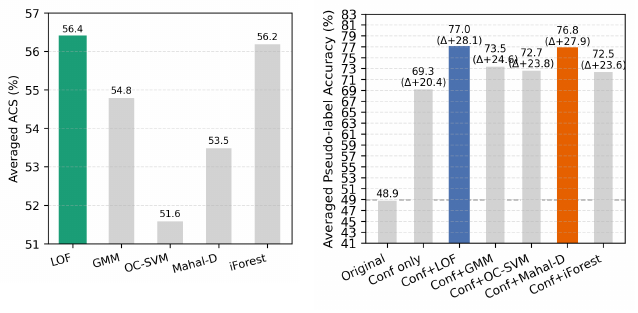}
  \caption{The left graph shows the averaged ACS scores for each outlier detection method combined with confidence filtering on MalwareDrift. The right graph
compares the averaged pseudo-label accuracy: original, with confidence filtering, with
confidence filtering + outlier detection. }\label{md_step2}
\end{figure}

Detailed accuracy and ACS scores are presented in Figure~\ref{md_step2} for MalwareDrift.

\end{document}